%% file: main.tex
\documentclass{article}

\usepackage{arxiv}

\usepackage{url,hyperref}

\include{pkghell.tex}
\include{cmdhell.tex}

\usepackage[utf8]{inputenc} 
\usepackage[T1]{fontenc}    

\usepackage{soul}

\newcommand{\citep}[1]{(\cite{#1})}

\title{\pkg{Moonshine.jl}: a \proglang{Julia} package for genome-scale model-based ancestral recombination graph inference}
\renewcommand{\shorttitle}{\pkg{Moonshine.jl}: ARG Inference}

\author{
 Patrick Fournier \\
  Département de Mathématiques\\
  Université du Québec à Montréal\\
  Montréal, Québec, H3C 3P8 \\
  \texttt{pf@patrickfournier.ca} \\
   \And
 Fabrice Larribe \\
  Département de Mathématiques\\
  Université du Québec à Montréal\\
  Montréal, Québec, H3C 3P8 \\
  \texttt{larribe.fabrice@uqam.ca} \\
}

\begin{document}
\maketitle
\begin{abstract}
\subimport{sections}{abstract.tex}
\end{abstract}

\keywords{Ancestral Recombination Graph \and ARG Inference \and Coalescent Theory \and Algorithms \and Julia}

\section{Introduction}
\subimport{sections}{introduction.tex}

\section{Tree Construction}\label{tree-construction}
\subimport{sections}{tree.tex}

\section{ARG Inference}
\subimport{sections}{arg.tex}

\section{Conclusion}
\subimport{sections}{conclusion.tex}

\section{Appendix}
\subsection{Simulations}\label{simulations}
\subimport{sections}{simulations.tex}

\subsection{Secretary Sampler}
\subimport{sections}{secretary-sampler.tex}

\bibliographystyle{unsrt}
\bibliography{refs}
\end{document}

%% file: pkghell.tex
\usepackage{import}

\usepackage{graphicx}

\usepackage{amsmath, amsfonts, amssymb}

\usepackage[dvipsnames]{xcolor}

\usepackage{tikz}
\usetikzlibrary{shapes.geometric}
\usetikzlibrary{shapes.arrows}
\usetikzlibrary{positioning}
\usepackage{memorygraphs}

\usepackage[linesnumbered,ruled,vlined]{algorithm2e}
\SetKwData{G}{g}
\SetKwData{U}{u}
\SetKwData{V}{v}
\SetKwData{Mrca}{mrca}
\SetKwData{Lr}{l\textsubscript{r}}
\SetKwData{Tmrca}{TgMRCA}
\SetKwData{Mrca}{mrca}
\SetKwData{Lr}{l\textsubscript{r}}
\SetKwData{Lc}{l\textsubscript{c}}
\SetKwData{Vr}{v\textsubscript{r}}
\SetKwData{Vc}{v\textsubscript{c}}
\SetKwData{Er}{e\textsubscript{r}}
\SetKwData{Ec}{e\textsubscript{c}}
\SetKwData{R}{r}
\SetKwFunction{Dst}{dst}
\SetKwFunction{Src}{src}
\SetKwFunction{Lat}{lat}
\SetKwFunction{AncestralInterval}{ancestral\_interval}
\SetKw{Continue}{continue}
\SetKw{Break}{break}
\SetKwFunction{Breakpoint}{breakpoint}
\SetKw{Goto}{go to}

\usepackage[nameinlink]{cleveref}
\crefname{subsection}{subsection}{subsections}
\Crefname{subsection}{Subsection}{Subsections}

\usepackage{multirow}
\usepackage{pifont}

\usepackage{paralist}

\usepackage{textcomp}

\usepackage{dirtytalk}

\usepackage{subcaption}

\usepackage{booktabs}

\usepackage{float}

\usepackage{amsthm}

\newtheorem{lemma}{Lemma}
\crefname{lemma}{lemma}{lemmas}
\Crefname{lemma}{Lemma}{Lemmas}

\usepackage{longtable}

\usepackage{setspace}

%% file: cmdhell.tex
\newcommand{\pkg}[1]{\texttt{#1}}
\newcommand{\proglang}[1]{\texttt{#1}}
\newcommand{\code}[1]{\texttt{#1}}

\newcommand{\given}{\,\vert\,}

\DeclareMathOperator{\ai}{\texttt{ancestral\_interval}}
\DeclareMathOperator{\idxtopos}{\texttt{idx\_to\_pos}}
\DeclareMathOperator{\postoidx}{\texttt{pos\_to\_idx}}

\definecolor{green}{HTML}{136F63}
\definecolor{red}{HTML}{C97064}
\definecolor{recarrow}{HTML}{D88DBE}

\renewcommand{\vec}[1]{\boldsymbol{#1}}
\DeclareMathOperator{\GF}{GF}
\DeclareMathOperator{\xor}{\oplus}

\newcommand{\Real}{\mathbb R}
\renewcommand{\emptyset}{\varnothing}
\DeclareMathOperator{\B}{\text{Beta}}

\renewcommand{\tilde}{\widetilde}

\newcommand{\edge}[2]{#1-#2}

\DeclareMathOperator*{\argmax}{arg\,max}

%% file: sections/abstract.tex
The ancestral recombination graph (ARG) is the model of choice in statistical genetics to model population ancestries. Software capable of inferring ARGs on a genome scale within a reasonable amount of time are now widely available for most practical use cases. While the inverse problem of inferring ancestries from a sample of haplotypes has seen major progress in the last decade, it does not enjoy the same level of advancement as its counterpart. Up until recently, even moderately sized samples could only be handled using heuristics. In recent years, the possibility of model-based inference for datasets closer to "real world" scenarios has become a reality, largely due to the development of threading-based algorithms. This article introduces \pkg{Moonshine.jl}, a \proglang{Julia} package that has the ability, among other things, to infer ARGs for samples of thousands of human haplotypes of sizes on the order of hundreds of megabases within a reasonable amount of time. On recent hardware, our package is able to infer an ARG for samples of densely haplotyped (over one marker/kilobase) human chromosomes of sizes up to 10000 in well under a day on data simulated by \pkg{msprime}. Scaling up simulation on a compute cluster is straightforward since each ARG is inferred independently using a single thread. While model-based, it does not resort to threading but rather places restrictions on probability distributions typically used in simulation software in order to enforce sample consistency. In addition to being efficient, a strong emphasis is placed on ease of use and integration into the biostatistical software ecosystem.

%% file: sections/introduction.tex
Coalescent theory is the framework of choice when it comes to inference from
genetic data \citep{Nordborg2007}. A brief introduction and some historical
notes on the subject can be found in \cite{Lewanski2024}. Although it originates
from the seminal papers of Kingman \citep{Kingman1982,Kingman1982a} and
Hudson \citep{Hudson1983}, it didn't really gain widespread use before the
advent of computer software capable of simulating the coalescent process with
recombination. \pkg{ms} \citep{Hudson1983}, a software written in \proglang{C},
was the first of its kind. It aims at generating a sample of genetic sequences
evolving under a Wright-Fisher (WF) model via a coalescent-with-recombination
(CWR) approximation. The WF model is known for its simplicity. Somewhat
surprisingly, simulating a sample of genetic sequences resulting from the
evolution of a population obeying it rapidly becomes a challenging task. This
is because it is necessary to keep track of every sequence at each of the
numerous discrete steps of the simulation procedure from the original population
to the desired sample. The coalescent process works in reverse: it starts
from the sample and generates coalescence events, which correspond to sets of
sequences finding a common ancestor. Consequently, it is only necessary to keep
track of haplotypes associated with these events. The remaining sequences are
deemed \emph{non-ancestral} for the sample and can be disregarded. In addition,
exponentially distributed coalescence times are substituted to discrete
generation counts. Sampling genetic sequences using the coalescent approximation
is a straightforward task.

What is not trivial, however, is relaxing the assumptions of the WF model
to allow recombination events. The result of simulations performed under
a model accounting for these events, along with, somewhat confusingly, the
statistical model itself, is known as the ancestral recombination graph
(ARG). To avoid ambiguity, we reserve ``ARG'' for the former and refer to the
statistical model through ``CWR''. While Hudson's \pkg{ms} is able to simulate
recombination events, it is not efficient enough to meet the requirements
of large-scale genomic data analysis. To reduce the computational burden
associated with simulations, an approximation to the CWR was developed by Niall
Cardin and Gil McVean~\citep{mcveanApproximatingCoalescentRecombination2005}.
Building on the work of \cite{Wiuf1999}, their idea is to approximate
the non-Markovian recombination process with a Markovian one called
the sequential Markov coalescent (SMC). This gave rise to a plethora
of so-called sequential simulators such as \pkg{MaCS} \citep{Chen2008},
\pkg{fastsimcoal} \citep{Excoffier2011}, \pkg{SC} \citep{Wang2014} and
\pkg{scrm} \citep{Staab2015}. While the sequential approach is not inherently
more efficient than the backward-in-time one, it is easier to approximate,
leading to substantial improvements in the capacity of CWR software.
Nonetheless, the idea of backward-in-time simulation was not abandoned, as
evidenced by \pkg{MSMS} \citep{Ewing2010} and \pkg{Discoal} \citep{Kern2016}.
\cite{messerSLiMSimulatingEvolution2013} even developed a forward-in-time
simulator able to handle non-WF scenarios. Some years later, the classical
backward-in-time CWR approach was brought back into the spotlight by
\pkg{msprime} \citep{Kelleher2016}. By reformulating Hudson's original
\pkg{ms} in terms of a data structure they called \emph{sparse trees}, their
\proglang{Python} package (with performance-critical procedures implemented in
\proglang{C}) has the capacity of simulating exactly from the coalescent with
recombination more efficiently than the sequential approximations available
at the time. Version 1.0 \citep{Baumdicker2021} brings even more features such
as the ability to simulate directly from a WF model. For all these reasons,
\pkg{msprime} is nowadays the \textit{de facto} reference for CWR and even ARG
simulation.

Programs mentioned above simulate genealogies with the goal of generating a
sample of genetic sequences. Throughout this paper, we refer to this process
as \emph{ARG simulation}; the ancestry is viewed as a sample point in the
probability space associated with the CWR. The focus is generally on inferring
parameters such as recombination/mutation rates or effective population
size. The distinctive characteristic of this class of algorithms is that the
distribution is parametrized by values derived from sequences of markers rather
than the markers themselves. This is to be contrasted with software designed
to solve the inverse problem of generating \emph{likely} ancestries, or even,
in some cases, a single ancestry directly from the markers. This is known as
\emph{ARG inference} since the ancestry itself is usually the main point of
interest. This choice of words is, however, somewhat misleading as so-called
inferred genealogies are not necessarily central to the analyses they are
involved in. They can be instrumental in the estimation of other parameters,
just like their simulated counterparts. It is true that their distribution is not
that of the CWR. That being said, this is not to say that alternative likelihood
functions enabling maximum likelihood estimation do not exist. Indeed, recent
work \citep{Bisschop2025} proposes formulations with applicability to broad
classes of ancestries in mind. While the problem these pieces of software solve is by
nature more computationally demanding, they provide major benefits in that the
ancestries they produce are generally more likely with respect to the sample
at hand. In particular, their usefulness cannot be overstated for methodologies
that aim at improving the estimation of parameters by treating the ancestry of
a sample as a latent variable such as \cite{larribeGeneMappingAncestral2002}
or \cite{Fournier2025}. Those require integrating these genealogies out, a task
involving, in practice, the ability to sample in high probability regions.

Early attempts at ARG inference include \pkg{recom} \citep{griffithsAncestralInferenceSamples1996a} and \pkg{Infs} \citep{fearnheadEstimatingRecombinationRates2001a}. More recently, the authors of \pkg{SC} also provides a modified version of their algorithm called \emph{SC-sample}, capable of solving this inverse problem by generating sample-consistent graphs. A graph is said to be consistent for a sample with respect to a mutation evolution model if it generates the sample under that model. In \cite{Wang2014}, they choose to use the popular infinite site model (ISM). Other software have been developed with the goal of inferring the ancestry of a sample. \pkg{ARGinfer} \citep{Mahmoudi2022} assumes a sample of sequences of binary markers evolving under the ISM. It aims at performing the inference probabilistically via a MCMC scheme allowing, for instance, computation of probability intervals for ARG-dependant quantities. Another noteworthy inference software is \pkg{ARGweaver} \cite{Rasmussen2014}, which implements an ancestry reconstruction algorithm based on the progressive integration of sequences to an existing ARG, an operation the authors call \emph{threading}. This approach forms the basis for recent methods such as \pkg{SINGER} \cite{Deng2025}. We should also mention \pkg{ARBORES} \cite{Heine2018}, a new take on the \emph{tree scan} methods \cite{Song2005} and \pkg{Espalier}, which implement its own original algorithm. These methods are designed to reconstruct ARGs according to a probabilistic model. For that reason, we refer to them as being \emph{model-based}.

A related problem is that of \emph{parsimony-based inference}, which consists of finding ARGs consistent with a sample using the minimum number of recombination events. This problem has been proved NP-hard \citep{wangPerfectPhylogeneticNetworks2001}. Nonetheless, attempts to solve it exactly and approximately date back to the early work of Hein \citep{heinReconstructingEvolutionSequences1990}, which was improved by \cite{Song2005}. One of the first widely available programs for parsimony-based inference is \pkg{Margarita} \cite{Minichiello2006}, which later inspired \pkg{ARG4WG} \cite{Nguyen2017} and \pkg{GAMARG} \cite{Thao2019}. \pkg{SHRUB} \cite{Song2005a} and \pkg{beagle} \cite{Lyngsoe2005} implement algorithms sharing similarities with that of \pkg{Margarita} although they were developed independently. The more recent \pkg{KwARG} has many features reminiscent of these two programs. Similar to \pkg{ARBORES}, \pkg{RENT} \cite{Wu2011} and its successor \pkg{RENT+} \cite{Mirzaei2016} are based on the tree scan method. \pkg{TMARG} implements two algorithms: one is related to \pkg{SHRUB}'s while the other derives from the notion of \textit{Steiner sequence}. \pkg{SARGE} \cite{Schaefer2021} implements a greedy algorithm built on top of the four gametes test \cite{Hudson1985}.

Some methods are partially founded on statistical frameworks, but employ heuristics at various levels. This is the case for \pkg{tsinfer} \cite{Kelleher2019}, \pkg{Relate} \cite{Speidel2019}, and \pkg{Threads} \cite{Gunnarsson2024}, which are based on the Li-Stephens model \cite{Li2003}. Specifically, \pkg{tsinfer} uses a bespoke consensus heuristic to infer ancestral haplotypes, \pkg{Relate} estimates the topology of marginal trees via hierarchical clustering and \pkg{Threads} uses a threading algorithm that involves matching haplotypes through a method based on the positional Burrows–Wheeler transform \cite{Durbin2014}. \pkg{ARG-needle} \cite{Zhang2023} relies on the Ascertained Sequentially Markovian Coalescent \cite{Palamara2018} for pairwise coalescence times estimation of subsets of sample haplotypes constructed by genotype hashing.

This paper introduces a \proglang{Julia} package called \pkg{Moonshine} implementing inference of sample-consistent ancestral recombination graphs. The approach is sequential, as it is based on iterative modification of the ARG. A sequence of operations are applied to a coalescent tree to ultimately make it consistent with a sample of haplotypes. The main advantage over back-in-time approaches is the possibility it offers users to specify various levels of approximation when generating ARGs. The available spectrum ranges from \emph{exact} simulation without any Markov assumption to first-order approximation \textit{à la} SMC. In any case, the object resulting from the ARG construction routine is the same, regardless of the level of approximation. It contains the whole graph, as well as meta-data such as vertex latitudes (number of generations from the sample), associated haplotypes, and intervals of ancestrality for edges, all available to users for subsequent analysis. In fact, \pkg{Moonshine} is designed for easy integration into data analysis workflows. In addition, it is possible to use it for inferring a set of ancestries consistent with a given sample of genetic sequences. It can be used interactively without sacrificing performance, thanks to \proglang{Julia}'s just-in-time compilation. It is also fully integrated with \proglang{Julia}'s ecosystem of graph theoretical packages. Although all sequential procedures are based on the same idea and, consequently, share many similarities, \pkg{Moonshine} implements its own original algorithm. Furthermore, since it is created with statistical inference in mind, \pkg{Moonshine} treats ARGs as random graphs. It is straightforward to evaluate ARG-related functions such as probability densities for the ARGs themselves or other random variables, such as phenotypes, conditional on an ARG. Implementation of custom functionalities is facilitated by a coherent type hierarchy and thorough documentation of abstract types and interfaces, making the extension of the package's various components as easy as possible. Finally, interoperability with \pkg{tskit} \citep{Kelleher2016, Wong2024} streamlines data management. Generating random samples directly from \pkg{tskit} objects is supported, and a convenience constructor for obtaining a sample from a simple genetic model using \pkg{msprime} \citep{Baumdicker2021} is provided. This is transparent to the end user, thanks to \pkg{Moonshine} being packaged with its own distribution of \pkg{msprime}. Additionally, the ARGs produced by \pkg{Moonshine} can be converted into \code{TreeSequence}s with a single function call. Conversely, \pkg{Moonshine} can be installed and used entirely from \proglang{Python} via the packages \pkg{JuliaPkg} and \pkg{JuliaCall}, respectively.

\pkg{Moonshine} shares with \pkg{SC-sample} the capacity of performing inference under various levels of approximation. To the best of our knowledge, these are the only two model-based inference methods implementing this functionality; \pkg{ARBORES}, \pkg{ARGWeaver}, \pkg{Singer}, and \pkg{Espalier} are based on SMC while \pkg{ARGinfer} does not allow for approximation. Allowing for approximate inference increases scalability by enabling a trade-off between biological realism and computational performance. On the other hand, \pkg{Moonshine} shares with \pkg{SC-Sample} and \pkg{ARGinfer} the ability to produce ancestries with complex correlation structures that may include type 2 recombination events (see \cref{unrestricted-rr}). Moreover, \pkg{Moonshine} supports multiple crossing over events (MCO), a unique feature among inference software. Unlike \pkg{ARBORES}, \pkg{ARGweaver}, \pkg{SINGER} and \pkg{ARGinfer}, \pkg{Moonshine} does not rely on MCMC and is thus fast and easy to use. No iterations are lost in burn-in periods or thinning intervals, and there is no need for convergence diagnostics or parameter tuning.

Our objective in developing \pkg{Moonshine} is not limited to creating
a realistic and convenient ARG inference software; performance is a major priority. We
present numerical experiments showing its potential for both coalescent tree
construction and ARG inference. Trees for large samples ($n = 10000$) of long
simulated haplotypes (250 Mbp) can be constructed in minutes at high resolution
(over one marker per kbp) using Hamming distance between sequences. In the
same scenarios, complete ancestries can be inferred in hours. Furthermore, our
algorithms are completely single-threaded, enabling us to increase sampling throughput by leveraging concurrency efficiently and easily by launching
multiple ARG inferrence instances in parallel. By being faster than other model-based alternatives, \pkg{Moonshine} represents a step towards enabling probabilistic inference of ancestries on the genome scale.

%% file: sections/tree.tex
Similar to other sequential procedures, the first step of our algorithm is to
construct an initial coalescent tree. Consistency with the first marker is not
assumed; ARG inference can be carried out even starting from a completely random
tree. The sole requirement is that it be a valid coalescent tree, i.e., a full
binary tree with a coherent set of latitudes for the vertices. The idea behind
this functionality is that it might be of interest to compare the performance
of a method under a model without recombination versus one that allows for such
events. It would make little sense in that context to give a special status to
a single marker, disregarding the remainder of the haplotypes. Consequently,
we give the user maximum flexibility when building coalescent trees, which may
be of interest both in their own right and as a stepping stone for constructing
more complex histories. As will be discussed later, \pkg{Moonshine} is packaged
with two haplotype metrics designed for tree building. It is straightforward for
the user to implement custom metrics.

Within our package, data structures representing genealogies are subtypes
of the \code{AbstractGenealogy} abstract type. The type of coalescent trees
is simply \code{Tree}. Given a sample (of type \code{Sample}) of phased and
polarized genetic sequences (of type \code{Sequence}), construction is controlled
by two parameters: the global mutation rate $\mu$ and a metric on haplotypes
$d$. Assuming a diploid population, sequences of length $l$ with $s$ markers,
a diploid effective population size $N_e$ and a constant per locus diploid mutation rate of
$\mu'$, the global mutation rate is $\mu = 4 N_e \mu' l$. These parameters are
either computed or explicitly passed to \code{Sample}'s constructor. As for the
metric, since \pkg{Moonshine} is compatible with binary markers exclusively,
a sample of size $n$ is an $n$-tuple $\vec H = (h_1, \cdots, h_n)$ of bit vectors of size $s$. Concretely, for each sequence $k$, we have
\begin{equation*}
  h_k = h_k^1 \cdots h_k^s
\end{equation*}
where $h_k^{\bullet} \in \{ 0, 1 \}$. Wild and derived alleles are
represented by 0 and 1 respectively as is standard in the literature. Let
$\xor$ denote addition modulo 2. Examples of useful metrics, also known as
\emph{distance functions}, include
\begin{align*}
  d_L(h_1, h_2) & = h_1^1 \xor h_2^1\\
  d_H(h_1, h_2) & = \sum_{i = 1}^s h_1^i \xor h_2^i\,.
\end{align*}
$d_L$ is the metric under which the distance between two sequences is zero
if and only if the state of their first marker is identical, 1 otherwise.
$d_H$ is the Hamming distance. Such discrete distances have a natural
biological interpretation as the number of mutations between (a subinterval
of) sequences. Arbitrary distances can be implemented by the user as subtypes
of \code{Distance}.

Detail-oriented readers might have noticed that the term ``metric'' is used
loosely, as the positivity axiom need not hold; the distance between two
distinct haplotypes may be 0. This is necessary to allow inference of trees
consistent for a single marker. Technically, the correct mathematical construct
is that of a pseudometric.

\subsection{Construction Algorithm}
Coalescence events are generated as follows: a vertex $v_a$ is choosen uniformly
among the set of \emph{live} vertices, that is, the vertices that have not
coalesced yet. Another vertex $v_b$ is selected conditional on $v_a$. The
probability $p_{a b}$ of selecting $v_b$ given $v_a$ is proportional to
\begin{equation}
  p_{a b} = \frac{\mu^{d_{a b}}}{\Gamma(d_{a b} + 1)}\label{eq:probmut}
\end{equation}
where $\Gamma$ is the gamma function, $d_{a b} = d(h_a, h_b)$ is the distance function with $h_a$ and
$h_b$ the haplotypes associated with $v_a$ and $v_b$ respectively. $p_{a b}$ is
an unnormalized Poisson probability, where $\Gamma$ is used instead of the usual
factorial function to allow non-integer distances. Both vertices then coalesce
into $v_c$ with associated haplotype $h_c = h_a \odot h_b$ where $\odot$ is
the Hadamard product. The shift in latitude is exponentially distributed with
rate parameter equal to the number of live vertices. The latitude of $v_c$ is
computed with respect to that of the previous event. This procedure replaces
$v_a$ and $v_b$ by $v_c$ in the set of live vertices. Repeating it $n - 1$ times
on a sample of $n$ haplotypes yields a coalescent tree.

The user has the possibility of biasing distance computation by specifying a
parameter $c_0 \in \Real_+ \cup \{ 0, \infty \}$ meaning that, in practice,
\begin{equation*}
  d_{a b} = c_0 d(h_a, h_b)\,.
\end{equation*}
$c_0$ can be used to reproduce the familiar behavior of other software. Setting $c_0 =
\infty$ and $d = d_L$, we obtain
\begin{equation*}
  d_{a b} = \begin{cases}
    0 &\mbox{if } h_a^0 = h_b^0\\
    \infty &\mbox{if } h_a^0 \neq h_b^0
  \end{cases}
\end{equation*}
resulting in
\begin{equation*}
  p_{a b} = \begin{cases}
    1 &\mbox{if } h_a^0 = h_b^0\\
    0 &\mbox{if } h_a^0 \neq h_b^0
  \end{cases}\,.
\end{equation*}
As a result, haplotypes with identical status at the first marker will be agregated, resulting in a tree consistent with that marker.

\begin{figure}
  \centering
  \includegraphics[width = \textwidth]{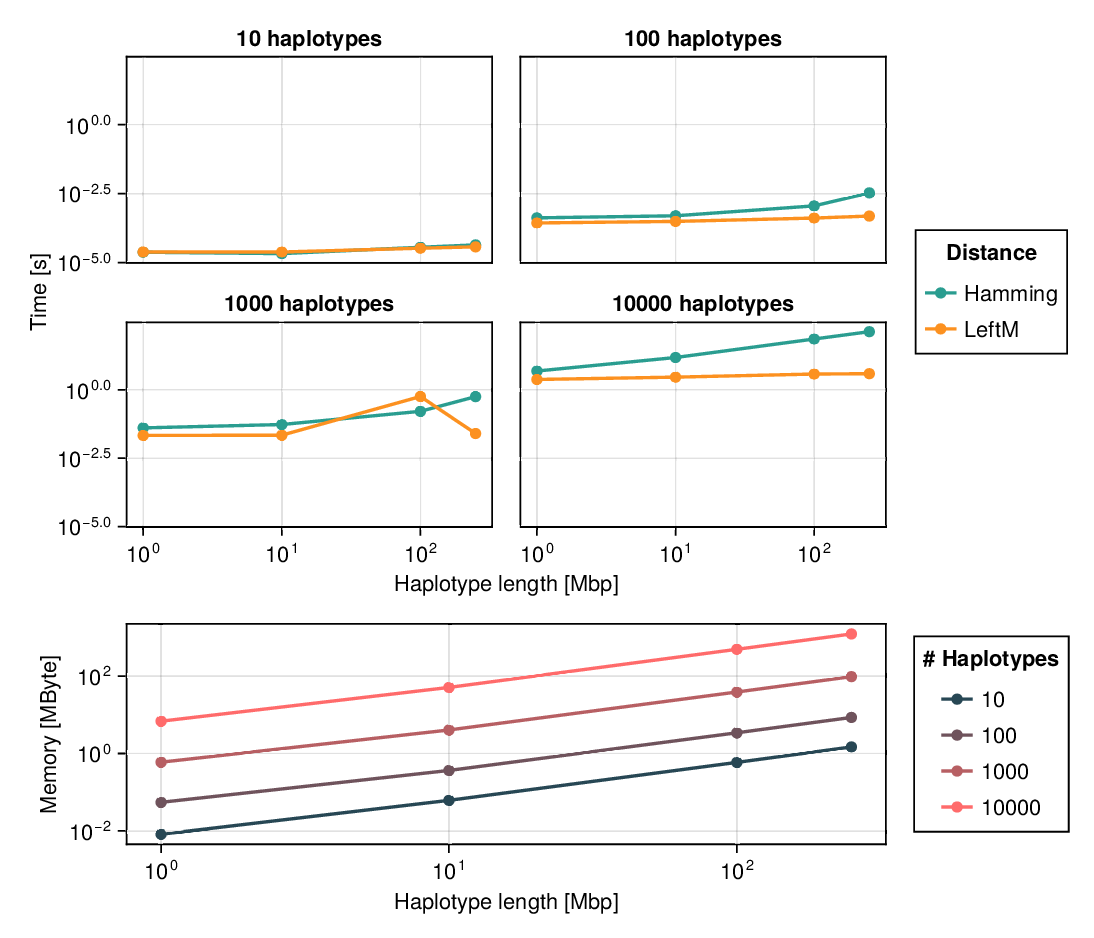}
  \caption{Time and memory needed for the construction of a coalescent tree as
  a function of the number of haplotypes, haplotype length, and distance
  function. Sampling is exact ($t_0 = 1$) and no bias is applied ($c_0 =
  1$). Construction with respect to either of the two distances yields identical
  memory usage since their computation does not involve memory allocation.
  Results are also available in
  \cref{table:tree-build-time-benchmark}.}\label{fig:tree-build-time-benchmark}
\end{figure}

\begin{figure}
  \centering
  \includegraphics[width = \textwidth]{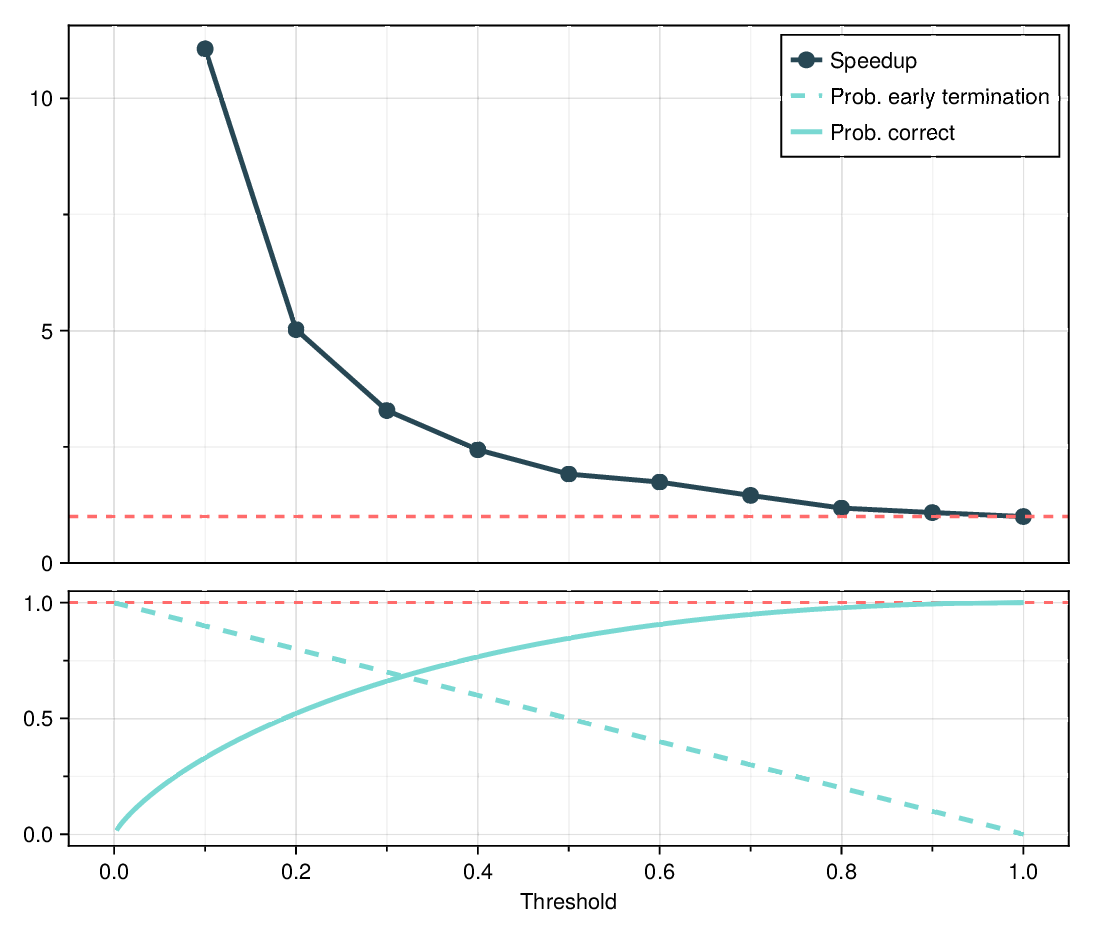}
  \caption{
    Speedup in tree construction, probability of selecting the correct sequence
    and probability of early termination as a function of the sampling
    threshold $t_0$. No bias is applied ($c_0 = 1$). The probability of early
    termination is $1 - t_0$. As the number of candidate vertices increases,
    the probability of sampling correctly from the target distribution converges
    to $t_0 (1 - \log t_0)$ (depicted here). The exact probability for a single
    run is given by \cref{lemma:secretary}. Results are also available in
    \cref{table:tree-sampling-threshold-benchmark}.
  }\label{fig:tree-sampling-threshold-benchmark}
\end{figure}

In practice, dealing with a ratio of such extreme quantities poses a numerical
challenge. For instance, if one wishes to use Hamming's distance, computation of
the normalizing constant becomes impossible even for relatively small values of
$n$ and $s$. We address these issues using two tricks. First, we compute $p_{a
b}$ via Stirling's approximation. On the logarithmic scale, we obtain
\begin{equation*}
  \log p_{a b} \approx d_{a b} (\log \mu - \log d_{a b} + 1)
\end{equation*}
which is already more manageable. Next, it would be ideal to refrain from reverting to the linear scale. Moreover, we would greatly benefit from avoiding the computation of the normalizing constant altogether. It turns out that this is exactly what the so-called \emph{Gumbel trick} \cite{gumbelStatisticalTheoryExtreme1954} is designed to do. The trick transforms sampling from the target distribution into an optimization problem. For a set of candidate vertices $b_1, \ldots, b_m$ and a sequence of iid standard Gumbel random variables $t_1, \ldots, t_m$, $\argmax_k \{ \log p_{a b_k} + t_k \}$ is distributed as a categorical random variable with the probability of category $k$ being equal to $p_{a b_k}$. This result gives us a more stable way of selecting the second coalescing vertex $v_b$, at the cost of increased computing time for drawing Gumbel random variables and finding the maximum of the sequence. In many cases, this impact should be minimal compared to the overall execution time, for instance when the tree is to be used for ARG inference. \Cref{fig:tree-build-time-benchmark} provides a rough estimate of time and memory usage for various scenarios. For cases where time considerations warrant a precision tradeoff, users can sample approximately from the target distribution via a scheme we call \emph{secretary sampling}, inspired by the famed \emph{secretary problem} \citep{fergusonWhoSolvedSecretary1989}. It revolves around giving the algorithm a chance of terminating before having traversed the complete set of candidate vertices. Its behavior is controlled by a user-determined threshold parameter $t_0 \in [0, 1]$. As illustrated in \cref{fig:tree-sampling-threshold-benchmark}, the probability of an early termination decreases with $t_0$. Note that although normalization is not required, the unnormalized probabilities associated with traversed vertices must be summed in order to evaluate the density of the resulting coalescent tree. This has to be done carefully as departure from the logarithmic scale is unavoidable. Details as well as the complete algorithm are presented in \cref{alg:tree-sampling}.

\begin{algorithm}
  \DontPrintSemicolon
  \linespread{1.25}\selectfont

  \SetKwFunction{P}{p}
  \SetKwFunction{Log}{log}
  \SetKwFunction{Exp}{exp}
  \SetKwFunction{Gumbel}{Gumbel}

  \SetKwData{Z}{z}
  \SetKwData{ZNew}{z'}
  \SetKwData{B}{b}
  \SetKwData{Va}{v\textsubscript{a}}
  \SetKwData{Vb}{v\textsubscript{b}}
  \SetKwData{LogP}{logp}
  \SetKwData{LogPNew}{logp'}
  \SetKwData{LogSP}{logc}
  \SetKwData{Idx}{idx}
  \SetKwData{K}{k}
  \SetKwData{PMin}{pmin}
  \SetKwData{PMax}{pmax}

  \SetKwArray{Live}{live}

  Fill array `\Live' with the live edges \;

  \BlankLine
  \While{$|\Live| > 1$}{
    Select \Va uniformly from \Live and remove it from \Live\;

    \BlankLine
    \tcp{$\log p_{a b} + \Gumbel$, $\log p_{a b}$, normalization constant and index of selected vertex respectively}
    Set $\Z \leftarrow \infty$,
    $\LogP \leftarrow 0$,
    $\LogSP \leftarrow 0$ and
    $\Idx \leftarrow 0$ \;

    \BlankLine
    Set $\K \leftarrow 0$ \;
    \While{$\K < |\Live|$}{
      Increment $\K \leftarrow \K + 1$ \;
      Set $\B \leftarrow \Live{\K}$, $\LogP \leftarrow \log{p_{a b}}$\;
      \BlankLine
      \If{$\LogP < \infty$}{
        Generate $\G \sim$ \Gumbel{0, 1}\;
        Set $\LogSP \leftarrow \LogP$, $\Z \leftarrow \LogSP + \G$ and $\Idx \leftarrow \K$ \;
        Break \;
      }
    }

    \BlankLine
    \While{$\K < |\Live|$}{
      Increment $\K \leftarrow \K + 1$ \;
      Set $\B \leftarrow \Live{\K}$, $\LogPNew \leftarrow \log{p_{a b}}$\;

      \BlankLine
      \If{$\LogPNew < \infty$}{
        \tcp{Update \LogSP accurately}
        Set $\PMin \leftarrow \min \{\LogPNew, \LogSP \}$,
        $\PMax \leftarrow \max \{\LogPNew, \LogSP \}$ \;
        Update $\LogSP \leftarrow \PMax + \log(1 + \exp(\PMin - \PMax))$ \;

        \BlankLine
        Generate $\G \sim$ \Gumbel{0, 1}\;
        Set $\ZNew \leftarrow \ \LogPNew + \G$ \;
        \If{$\ZNew \geq \Z$}{
          Set $\LogP \leftarrow \LogPNew$, $\Z \leftarrow \ZNew$, $\Idx \leftarrow \K$ \;
          \If{$\K >$ threshold}{
            Break \;
          }
        }
      }
    }

    \BlankLine
    \If(\tcp*[h]{All probabilities were infinite}){$\Idx = 0$}{
      Select \Idx uniformly from $1, \ldots, |\Live|$ \;
      Set $\LogP \leftarrow 0$, $\LogSP \leftarrow \log |\Live|$ \;
    }

    \BlankLine
    Set $\B \leftarrow \Live{\Idx}$, $\Vc \leftarrow 2 n - |\Live|$ \;
    Remove \Vb from \Live, coalesce \Va and \Vb into \Vc and add \Vc to \Live \;

    Add $\LogP - \LogSP - \log(1 + |\Live|)$ to tree's log density \;
  }

  \caption{Tree Construction}\label{alg:tree-sampling}
\end{algorithm}

%% file: sections/arg.tex
In \pkg{Moonshine}, ancestral recombination graphs are instances of the
\code{ARG} type, a subtype of \code{AbstractGenealogy}. Since they represent
a more realistic model for the ancestry of a sample subject to recombination,
ARGs can be viewed as improved versions of coalescent trees. \code{ARG}s
are constructed sequentially from a \code{Tree} by iteratively generating
recombination events until an ancestry consistent with the sample is reached.
As is common, consistency is defined through the ISM: a genealogy is consistent
with a sample if the number of mutations per marker is at most one. In a context
where the recombination rate is several orders of magnitude higher than the
mutation rate, a consistent ARG is typically a more realistic genealogy than any
other kind of inconsistent ancestry. \pkg{Moonshine} is consequently very well
suited to working with single nucleotide polymorphisms (SNPs), which have a low
mutation rate.

Ancestries are modified by recombination events, which partitions ancestral
material into two subintervals: that to the left of an associated point, called
a \emph{breakpoint}, and the material to the right. From a graph-theoretical
perspective, a recombination event is generally represented by a vertex of
degree 3 with a single child and two parents. As an example, imagine that the
child edge of a recombination vertex is ancestral for an interval $I$ and that
the associated breakpoint is $b$. In that case, one of the parental edges,
generally referred to as the left edge, is ancestral for $[0, b) \cap I$ while
the other (right) edge is ancestral for $[b, \infty) \cap I$. Although we
used the term ``interval'', $I$ can actually be a union of intervals. We will
continue to use this terminology when the distinction between the two concepts
is not relevant.

\begin{figure} 
  \centering
  \begin{subfigure}[t]{0.3\textwidth}
    \centering
    \begin{tikzpicture}
      \node (figure) {
        \includegraphics[width = 0.875\columnwidth]{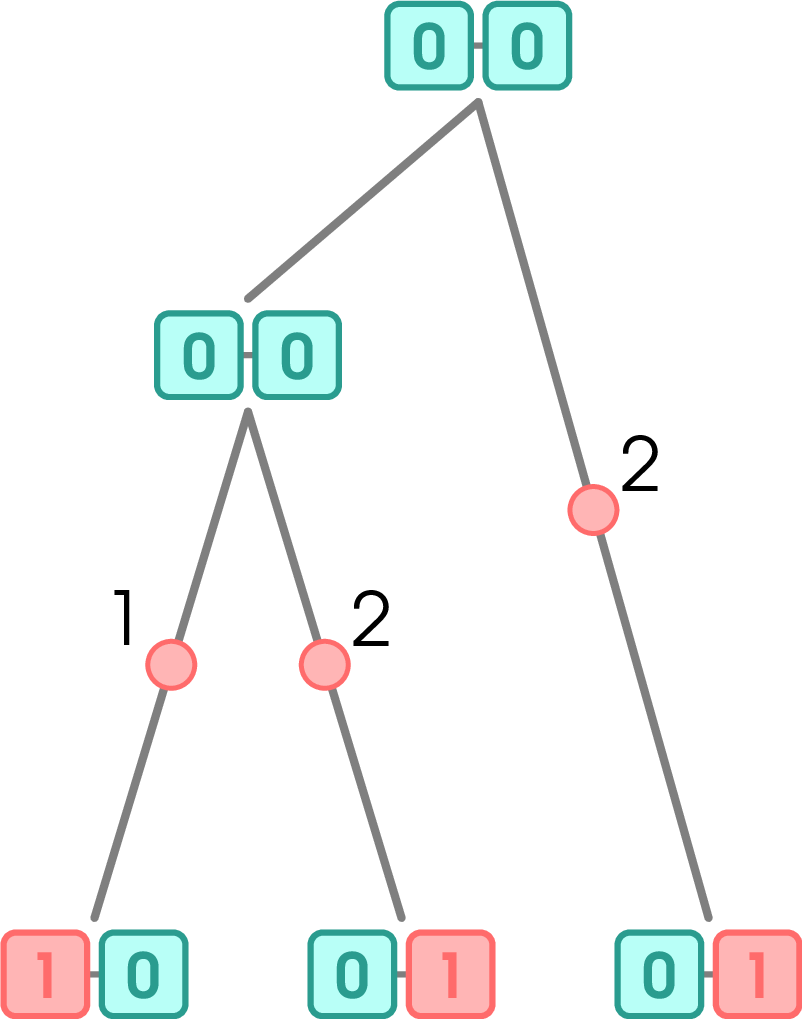}
      };
      \node[isosceles triangle,
            draw = recarrow,
            inner sep = 0pt,
            rotate = 90,
            opacity = 0,
            minimum size = 2mm] at (-1.25, -3.05) {};
    \end{tikzpicture}
    \caption{Initial graph.}\label{fig:recombination-above-below}
  \end{subfigure}
  \hfill
  \begin{subfigure}[t]{0.3\textwidth}
    \centering
    \begin{tikzpicture}
      \node (figure) {
        \includegraphics[width = \columnwidth]{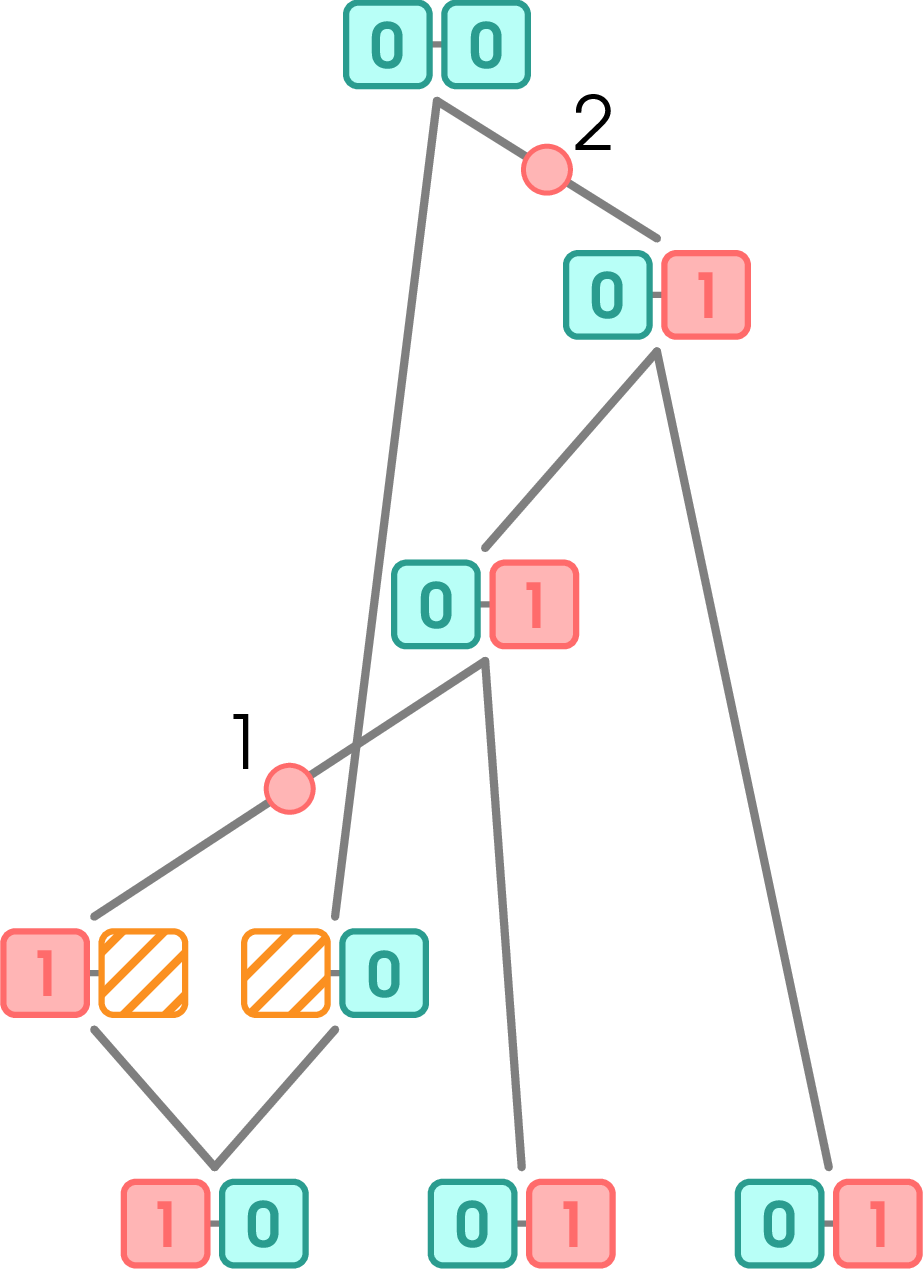}
      };
      \node[isosceles triangle,
            draw = recarrow,
            inner sep = 0pt,
            rotate = 90,
            fill = recarrow,
            minimum size = 2mm] at (-1.25, -3.7) {};
    \end{tikzpicture}
    \caption{Recoalescence above the gMRCA.}\label{fig:recombination-above}
  \end{subfigure}
  \hfill
  \begin{subfigure}[t]{0.3\textwidth}
    \centering
    \begin{tikzpicture}
      \node (figure) {
        \includegraphics[width = 0.875\columnwidth]{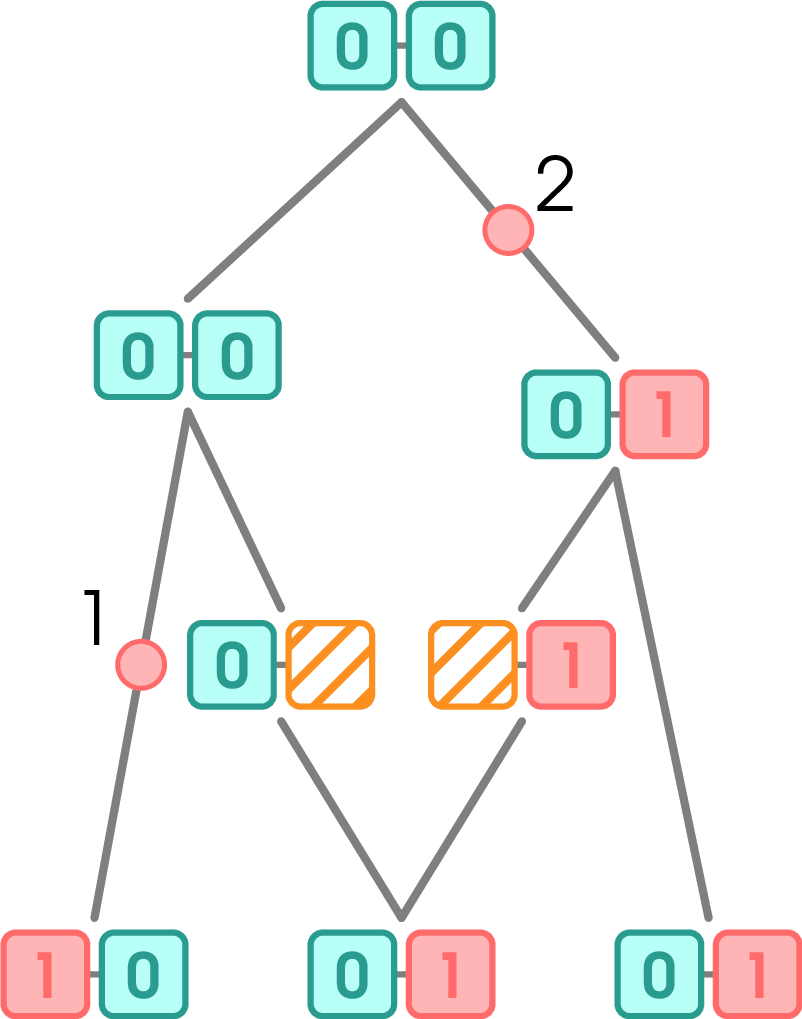}
      };
      \node[isosceles triangle,
            draw = recarrow,
            inner sep = 0pt,
            rotate = 90,
            fill = recarrow,
            minimum size = 2mm] at (0, -3.05) {};
    \end{tikzpicture}
    \caption{Recoalescence below the gMRCA.}\label{fig:recombination-below}
  \end{subfigure}

  \caption{Two types of recombination events. Blue and red boxes represent SNPs with the wild and derived alleles, respectively. Non-ancestral material is represented by yellow boxes. Mutation edges are marked with red dots and numbers indicating the mutating markers. Pink arrows indicate the breakpoints. When recoalescence occurs \emph{below} the MRCA of the sample (gMRCA), the recoalescence vertex is ``inserted'' by deleting an edge from the graph and replacing it with two new edges incident to the recoalescence vertex; this is illustrated in \cref{fig:recombination-below}. By contrast, no deletion is required when recoalescence occurs \emph{above} the gMRCA; as illustrated in \cref{fig:recombination-above}, the recoalescence vertex simply becomes the new gMRCA.}\label{fig:recombinations}
\end{figure}

Recombination vertices are ``added'', figuratively speaking, by deleting an
edge from the graph and connecting the new vertex with both endpoints of the
removed edge. The edge connected to the child vertex is the recombination
vertex's child edge, and the other is its left edge. This procedure leaves
the right edge floating. Every recombination is immediately followed by
the coalescence of the right edge with the graph. As it is carried out by a
different algorithm than the coalescence of two dangling vertices encountered
in temporal algorithms or, more trivially, when inferring a coalescent tree, we
call those \emph{recoalescence} events even though they result in an additional
coalescence vertex as well. Coalescence and recombination vertices are in many
regards mirror images of each other. A coalescence vertex has two child edges
and one parental edge. If the child edges are ancestral for two intervals
$I_l$ and $I_r$, then so is the parental edge for $I_l \cup I_r$. The standard
recoalescence procedure begins again by deleting an edge, followed by connecting
its incident vertices with the recoalescence vertex. The remaining child edge
is then connected to the recombination vertex's right edge, concluding the
procedure. Recoalescence can also occur without edge deletion. In that case, the
coalescence vertex becomes the new root of the graph and lacks a parental edge.
It is connected downstream to the previous root and the recombination vertex. Both types of recoalescence, with and without edge deletion, are illustrated in \cref{fig:recombinations}.

The type of a recoalescence event depends on its latitude, denoted $l_c$,
which itself depends on the recombination's latitude $l_r$. The root of an
ARG corresponds to the sample's most recent common ancestor (MRCA), sometimes
called the \emph{grand} MRCA (gMRCA). Its latitude is the time to the
gMRCA (TgMRCA). Let $v_r$ and $v_c$ be the recombination and recoalescence
vertices and $\edge{s_r}{d_r}$ and $\edge{s_c}{d_c}$ the edges deleted in the
recombination-and-recoalescence (RR) steps. \Cref{alg:recombination} summarizes
the RR procedure.

\begin{algorithm}
  \DontPrintSemicolon

  \SetKwData{Sr}{s\textsubscript{r}}
  \SetKwData{Dr}{d\textsubscript{r}}
  \SetKwData{Sc}{s\textsubscript{c}}
  \SetKwData{Dc}{d\textsubscript{c}}
  \SetKwData{G}{G}

  Add two vertices \Vr and \Vc to \G \;
  Delete recombination edge \Sr ~--- \Dr \;
  Add edges \Sr ~--- \Vr and \Vr ~--- \Dr \;
  \eIf{$\Lc \leq \Tmrca$}{
    Delete recoalescence edge \Sc ~--- \Dc \;
    Add edges \Sc ~--- \Vc and \Vc ~--- \Dc \;
  }{
    Add edge \Vc ~--- \Mrca \;
  }
  Add edge \Vc ~--- \Vr \;

  \caption{Recombination-and-recoalescence}\label{alg:recombination}
\end{algorithm}

\subsection{Unrestricted Recombination-and-Recoalescence events}\label{unrestricted-rr}
\pkg{Moonshine} has the capability to generate two kinds of RR events: restricted
and unrestricted. Unrestricted events, the subject of this section, have a
distribution designed to closely match that of the CWR. Restricted recombination
events are designed to reduce the total number of mutation events on an ARG;
these will be discussed at length in the next section.

\begin{figure*} 
  \centering
  \begin{subfigure}[t]{0.3\textwidth}
    \centering
    \begin{tikzpicture}
      \node (figure) {
        \includegraphics[width = \columnwidth]{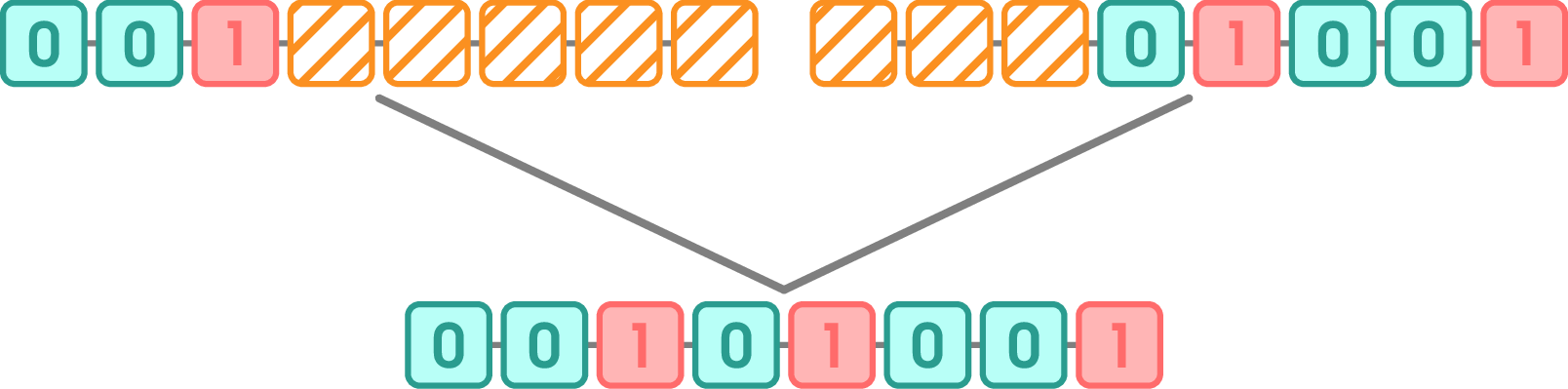}
      };
      \node[isosceles triangle,
            draw = recarrow,
            below = of figure,
            inner sep = 0pt,
            rotate = 90,
            fill = recarrow,
            minimum size = 2mm] at (-0.3, 0.1) {};
      \end{tikzpicture}
    \caption{Type 1}\label{fig:recombinations-types-1}
  \end{subfigure}
  \hspace{3ex}
  \begin{subfigure}[t]{0.3\textwidth}
    \centering
    \begin{tikzpicture}
      \node (figure) {
        \includegraphics[width = \columnwidth]{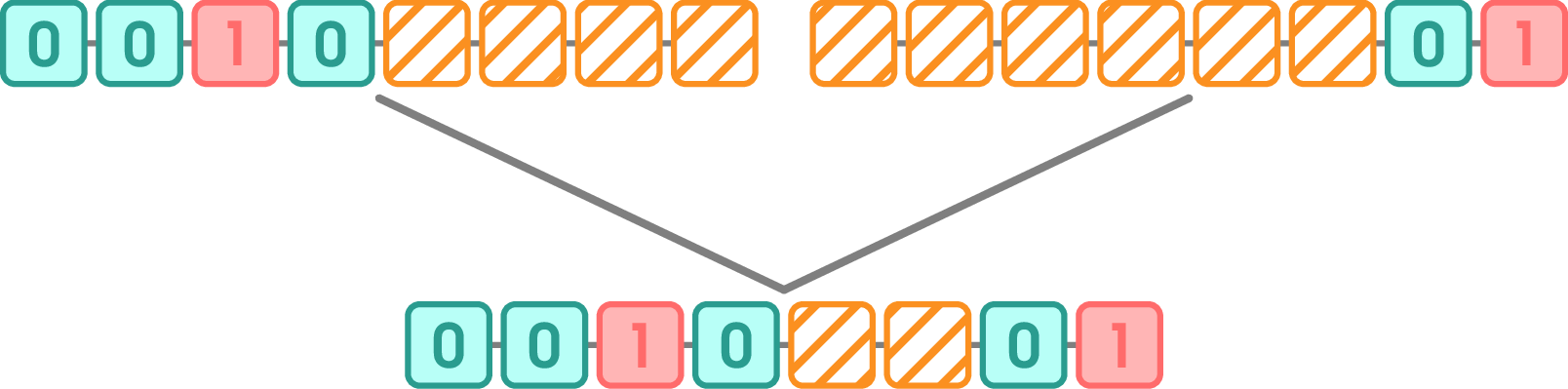}
      };
      \node[isosceles triangle,
            draw = recarrow,
            below = of figure,
            inner sep = 0pt,
            rotate = 90,
            fill = recarrow,
            minimum size = 2mm] at (0.2, 0.1) {};
      \end{tikzpicture}
    \caption{Type 2}\label{fig:recombinations-types-2}
  \end{subfigure}

  \vspace{3em}
  \begin{subfigure}[t]{0.3\textwidth}
    \centering
    \begin{tikzpicture}
      \node (figure) {
        \includegraphics[width = \columnwidth]{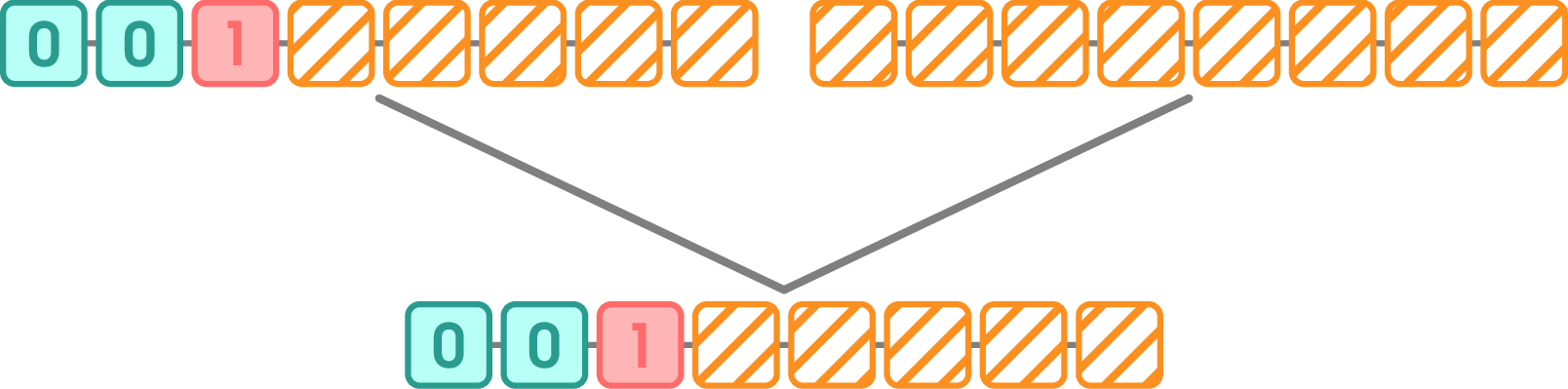}
      };
      \node[isosceles triangle,
            draw = recarrow,
            below = of figure,
            inner sep = 0pt,
            rotate = 90,
            fill = recarrow,
            minimum size = 2mm] at (0.55, 0.1) {};
      \end{tikzpicture}
    \caption{Type 3}\label{fig:recombinations-types-3}
  \end{subfigure}
  \hfill
  \begin{subfigure}[t]{0.3\textwidth}
    \centering
    \begin{tikzpicture}
      \node (figure) {
        \includegraphics[width = \columnwidth]{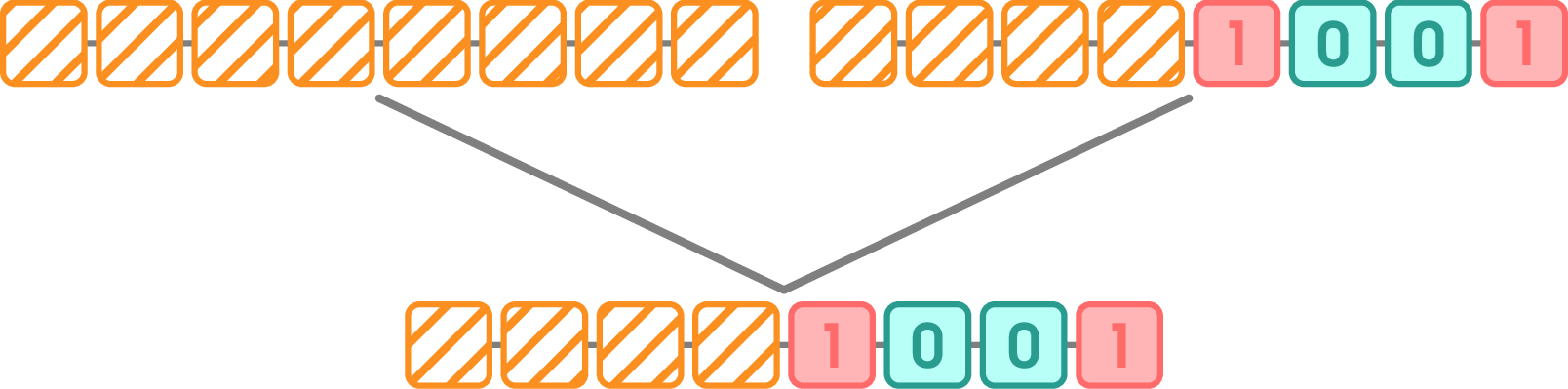}
      };
      \node[isosceles triangle,
            draw = recarrow,
            below = of figure,
            inner sep = 0pt,
            rotate = 90,
            fill = recarrow,
            minimum size = 2mm] at (-0.5, 0.1) {};
      \end{tikzpicture}
    \caption{Type 4}\label{fig:recombinations-types-4}
  \end{subfigure}
  \hfill
  \begin{subfigure}[t]{0.3\textwidth}
    \centering
    \begin{tikzpicture}
      \node (figure) {
        \includegraphics[width = \columnwidth]{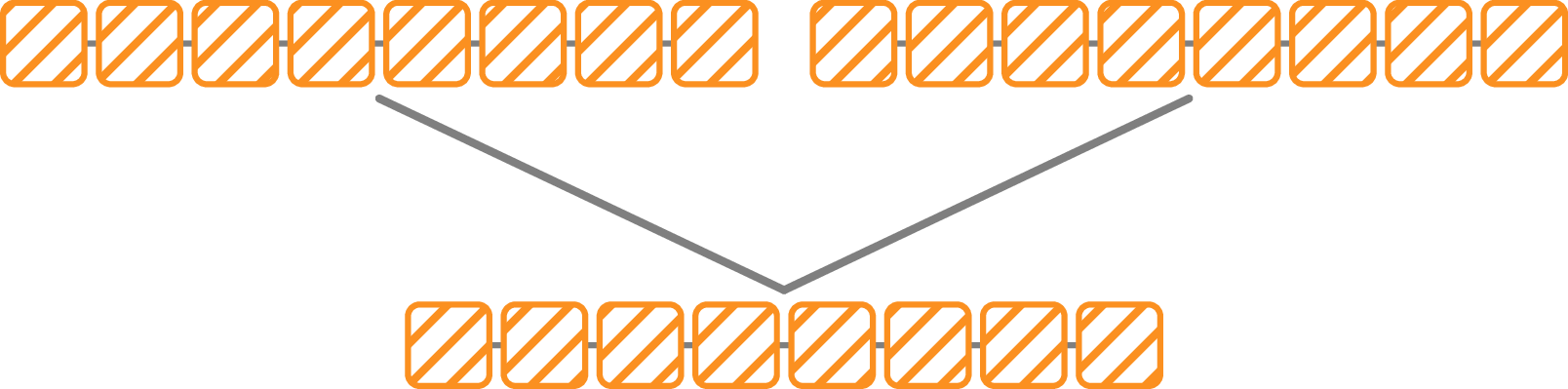}
      };
      \node[isosceles triangle,
            draw = recarrow,
            below = of figure,
            inner sep = 0pt,
            rotate = 90,
            fill = recarrow,
            minimum size = 2mm] at (-0.8, 0.1) {};
      \end{tikzpicture}
    \caption{Type 5}\label{fig:recombinations-types-5}
  \end{subfigure}

  \caption{Types of recombination events. Pink arrows indicate the positions of breakpoints.}\label{fig:recombinations-types}
\end{figure*}

Our package exports a method for generating an arbitrary number of unrestricted
recombination events. Standard theory~\cite{Wiuf1999} models the positions
(on sequences) and locations (on ancestries) of recombination events as a
Poisson point process (PPP). Conditional on their number, both locations and
positions are distributed uniformly. When applied directly, this method has the
drawback of generating sequences devoided of material ancestral for the sample
at hand. Recombination events can be classified depending on whether ancestral
material is present on both sides. If it is, the breakpoint can be positioned
in ancestral or non-ancestral material. These are referred to as \emph{type 1}
and \emph{type 2} events respectively and exclusively create haplotypes having
ancestral material. Events positioned such that only material to their left
or right side is ancestral are classified as \emph{type 3} and \emph{type 4}
respectively. Finally, events occurring in entirely non-ancestral sequences
are classified as \emph{type 5}. Examples for each type of event are provided in \cref{fig:recombinations-types}. It is desirable for an algorithm to only
generate the first two types of recombination events since the other ones do
not contribute to the structure of the sample. Our method follows this approach.
We start by drawing a recombination edge with probability proportional to its
length. The location of the event is distributed conditional on the selected
edge. Then, we choose a position, also known as a \emph{breakpoint}, uniformly
on the mathematical closure of the set of intervals for which the recombination
edge is ancestral. This strategy enforces the existence of ancestral material
on both sides of the breakpoint and avoids recombination events of type 3, 4 and
5. Additionally, since the closure of ancestral intervals is not, in general,
equal to the intervals themselves (it may contain ``holes'' of non-ancestral
material), type 2 events are possible.

Selecting a conditional distribution for the location of recombination events
on branches requires careful consideration. Theory dictates that it should be
uniform. However, this results in a problematic frequency of locations close
to the branches' endpoints, contributing to the \emph{short branches issue}
discussed in \cref{recombination-recoalescence}. Consequently, we use the
location-scale family associated with the $\text{Beta}(2, 2)$ distribution
instead. This method preserves the symmetry of the uniform distribution while
favoring locations closer to the center of branches, achieving sufficient
reduction in the number of short branches to prevent numerical error.

The recoalescence process follows standard theory. The latitude is
distributed as an inhomogeneous Poisson process with rate equal to the
number of branches. The usual time-scale transformation strategy (see
\cref{recombination-recoalescence}) is implemented. A recoalescence
edge is selected uniformly, conditional on the latitude. The process
of generating a single unrestricted recombination event is summarized in
\cref{alg:unrestricted-recombination}.

\begin{algorithm}
  \DontPrintSemicolon

  \SetKwData{Lr}{l\textsubscript{r}}
  \SetKwData{Lc}{l\textsubscript{c}}
  \SetKwFunction{Dst}{ldst}
  \SetKwFunction{Src}{lsrc}

  Select a recombination edge in $\Er \in E$ with probability proportional to
  its length \;

  Choose a breakpoint $\R \in (r_L, r_U)$ where $r_L$ and $r_U$ are the
  leftmost and rightmost positions for which \Er is ancestral
  \;\label{unconstrained-recombination-bp}

  Generate a recombination latitude \Lr distributed as $\Dst{\Er} + (\Src{\Er} -
  \Dst{\Er}) \B(2, 2)$ where \Src{\Er} and \Dst{\Er} are the latitudes of the
  highest and lowest vertex adjacent to \Er respectively
  \;\label{unconstrained-recombination-rlat}

  Generate a recoalescence latitude \Lc distributed as an inhomogeneous PPP with
  rate equal to the number of branches via time-scale transformation \;

  Select a recoalescence edge \Ec uniformly among the edges at latitude \Lc \;

  Apply \cref{alg:recombination} \;

  \caption{Unrestricted Recombination}\label{alg:unrestricted-recombination}
\end{algorithm}

As a sidenote, since a function allowing users to add arbitrary recombination
events to a graph is exposed, \pkg{Moonshine} can easily be used to simulate
ancestries. We do not recommend doing so, however, especially not using the
built-in types \code{Tree} and \code{Arg}, which are designed to implement ARG
inference. Instances of these types store information irrelevant in a simulation
context which greatly hinders performance.

\subsection{Restricted Recombination-and-Recoalescence Events}
\pkg{Moonshine} can be used for ARG inference due to its ability to
transform an inconsistent graph into a consistent one by generating a sequence
of RR events. A typical run starts with the construction of a coalescent tree and is
followed by a sequential sweep made up of the following broad steps:
\begin{enumerate}
  \item Start at the leftmost position;
  \item Find the next position that is inconsistent with the
        sample;\label{it:inconsistency-detection}
  \item Generate a restricted recombination event;
  \item Repeat until the rightmost position is reached.
\end{enumerate}

To ensure our graphs' distribution closely resembles that of the CWR, our aim,
alongside computational efficiency, is to impose as few constraints as possible
on restricted recombination events. The probability distributions involved are
similar to their unrestricted counterparts, except for their support, which we
attempt to reduce only as much as is necessary to enforce consistency of the
final ARG.

\subsubsection{SIMD-Accelerated Minimum Mutation Number Algorithm}
Step \ref{it:inconsistency-detection} requires an efficient inconsistency
detection algorithm. Specifically, given a starting position, it must be able to
find the closest marker to its right that mutates more than once in its current
marginal tree. Furthermore, efficiency considerations demand that the set of
edges on which these mutation events occur, which we refer to as \emph{mutation
edges}, is computed simultaneously. We refer to algorithms fulfilling these
two requirements as \emph{minimum mutation number} (MMN) algorithms. Mutation
edge identification is at the core of Moonshine's ARG inference procedure
and it should come at no surprise that a lot of time and effort went into its
optimization. It is key to enabling inference on real-world sized datasets.
We begin this section by giving a general description of the procedure before
delving into implementation specifics.

Our algorithm leverages the parallel introduced in \cref{tree-construction}
between sequences of $s$ biallelic markers and $s$-vectors of $\GF(2)$. In
addition to $\oplus$ defined earlier, denote the multiplication on $\GF(2)$,
which is identical to the one for natural numbers, by $\otimes$. Within that
framework, both binary operations can be given a biological interpretation. Let
$h_1$ and $h_2$ be two vectors of $\GF(2)^s$ and let $h = h_1 \oplus h_2$ where
$\oplus$ is applied elementwise. Since the characteristic of $\GF(2)$ is 2 (i.e.
$1 \oplus 1 = 0$), any non-zero element $h^{(0)}$ of $h$ results from $h_1^{(0)}
\neq h_2^{(0)}$. This, in turn, indicates an odd number of mutations occurring
between the two sequences at that marker. Assuming a mutation model disallowing
back mutations, this simplifies to exactly one mutation event.

Interpretation of $\otimes$ is slightly more convoluted. Let $v$ be an internal
vertex and $v_1, \ldots, v_m$ the subset of leaves having $v$ as an ancestor.
Denote the corresponding sequences by $h_v, h_{v_1}, \ldots, h_{v_m}$. Assuming
no back mutations,
\begin{equation*}
  h_v = h_{v_1} \otimes \cdots \otimes h_{v_m}
\end{equation*}
where, again, $\otimes$ is applied elementwise. This is easily seen to be the
case for a vertex $v$ having two leaves $v_1$ and $v_2$ as children. The only
way $h_v^k = 1$ is if $h_{v_1}^k = h_{v_2}^k = 1$ since at most one $k$-mutation
could have occurred on $\edge{v}{v_1}$ and $\edge{v}{v_2}$. For the same reason,
if $h_v^k = 0$, then either $h_{v_1}^k = h_{v_2}^k = 0$, in which case no
mutation occurred, or $h_{v_1}^k = h_{v_2}^k \oplus 1$, in which case a single
one occurred. This explanation can be extended to ``deeper'' vertices by moving
upward, labelling internal vertices encountered along the way.

The correspondence between $\GF(2)$ and the evolution of sequences of
biallelic markers under a model precluding back mutations can be exploited for
computational efficiency. In practice, for large samples, our implementation
reduces the time needed to compute mutation edges to about 15\% of total
execution on recent hardware, and this is without resorting to any form of
multithreading. Nearly all of the time spent in the MMN algorithm is devoted
to graph traversal. This level of performance is made possible by the following
data structure: A sequence $h = h^1 \cdots h^s$ can be associated to the number
$\tilde h$ whose binary representation has its first $s$ least significant
digits corresponding to $h$. In order to make this representation unique, we
assume that the other digits of $\tilde h$ are 0. This corresponds to
\begin{equation*}
  \tilde h = \sum_{k = 1}^s 2^k h^k \,.
\end{equation*}
When context is unambiguous, we shall drop the tilde and denote by $h$ either
a haplotype, the corresponding vector or its integer representation. Treating
haplotypes as integers is very convenient as operations $\oplus$ and $\otimes$
described above correspond to the bitwise exclusive disjunction (\code{XOR})
and conjunction (\code{AND}). For that reason, we denote those by $\oplus$ and
$\otimes$ as well. This interpretation allows for very efficient handling of
sequences since these are single instructions in computer processors: we cannot make mutation finding any simpler or faster with respect to computational hardware. Given an edge $v_1-v_2$ ancestral for an
interval $\omega$, we can find all of its mutations via \cref{alg:mmn-edge}:

\begin{algorithm}
  \DontPrintSemicolon

  Compute a $\otimes$-mask $m_{\omega}$ masking markers outside $\omega$ to 0 \;
  Compute $h = (h_{v1} \oplus h_{v_2}) \otimes m_{\omega}$ \;
  Find the indices of set bits in $h$ \;

  \caption{Mutation Events on an Edge}\label{alg:mmn-edge}
\end{algorithm}

Modern computers generally read data from memory in chunks of 64 bits and
can apply elementary bitwise operations such as \code{XOR} and \code{AND} to
such chunks in a single cycle. This is readily implemented and results in the
concurrent assessment of 64 markers on two haplotypes on a single physical
core. This is good, but we can do even better. On recent hardware, some
operations can be applied to chunks larger than 64 bits, a capability enabled
by so-called single instruction, multiple data (SIMD) instructions. At the
upper end of the spectrum are processors supporting the AVX-512 instruction set,
which can operate on 512 bits simultaneously. Such operations are said to be
``vectorized'' in reference to them being applied to multiple 64 bits ``scalars''
at once. While not all regular CPU operations have a vectorized counterpart,
integer \code{AND} and \code{XOR}, the only two operations required for mutation
detection, are part of the AVX-512F extension available on any AVX-512 compliant
processor.

That being said, it would generally not be very efficient to compare chunks of
512 markers from 2 different haplotypes simultaneously, as the next mutating
marker is likely near the last recombination breakpoint. Instead, we process
haplotypes in chunks of 8 markers. On an AVX-512 enabled architecture, this
results in up to 64 haplotypes being compared at 8 positions concurrently
without multithreading. In fact, this is so efficient that tested multithreaded
versions were actually slower than the single-threaded one. Our algorithm is
implemented in \pkg{Moonshine} using \pkg{SIMD.jl}~\citep{Schnetter2025} which
ensures portability across architectures. A custom fallback is implemented to allow use on hardware that does not support AVX.

Computing the number of mutations by batches of 8 markers naturally leads to optimizations of the graph traversal procedure. The main one is based on the observation that, assuming no back mutation, a given marker cannot have the derived allele for a haplotype if one of its descendants has the wild allele. This is the biological version of 0 being the absorbing element of $\otimes$. By performing traversal in a bottom-up fashion starting at the leaves, we allow for early termination when encountering a vertex associated with a chunk of zeros. The probability of early termination increases as the size of the chunk of markers decreases, further justifying our decision to work with byte-sized chunks. The complete MMN procedure is given in \cref{alg:mmn}. Storing traversed edges and associated chunks before applying \cref{alg:mmn-edge} (line \ref{line:mmn-store}) is wasteful but necessary to exploit SIMD parallelism. Line \ref{line:mmn-ctz} assumes that markers are encoded right-to-left within chunks (e.g., the sequence 111000 is stored as \code{000111}). On modern architectures, \code{trailing\_zeros} essentially correspond to a single processor instruction on modern hardware and is consequently very fast. \code{idx\_to\_pos} (line \ref{line:mmn-idxtopos}) is the function that computes the position of a marker from its index; see \cref{markers-positions} for details. The chunk size of 1 byte (8 bits), assumed throughout \cref{alg:mmn}, is a compile-time constant for performance reasons. Although we believe it to be the best choice in most circumstances, it can easily be configured via Julia's preferences mechanism.

\begin{figure*}
  \centering
  \begin{tikzpicture}[memory graph]
    \node[arity = 2, minimum height = 4em] (srcs) {
      $\overbrace{\underbrace{10110101}_{\text{chunk } 1} \cdots \underbrace{01000100}_{\text{chunk } 64}}^{64 \times 8 = 512 \text{ markers}}$
      \arg{1} \dots
      \arg{2} $00101001 \cdots 00011001$};
    \node[left = of srcs] (srcs-label) {Parents};

    \node[arity = 2, below = 0.5em of srcs] (dsts) {
      $11110101 \cdots 01110100$
      \arg{1} \dots
      \arg{2} $00111101 \cdots 00111111$};
    \node[left = of dsts] (dsts-label) {Children};

    \node[
      single arrow,
      anchor = west,
      yshift = -1em,
      rotate = 270,
      draw = green,
      fill = green,
      fill opacity = 0.75,
      minimum width = 10pt,
      single arrow head extend=3pt,
      minimum height=10mm
    ] (arrow-xor) at (dsts.south) {};
    \node [right = 0.25em of arrow-xor.north] (xor-label) {\scalebox{2}{$\oplus$}};

    \node[arity = 2, below = 1em of arrow-xor.east] (xors) {
      $01000000 \cdots 00110000$
      \arg{1} \dots
      \arg{2} $00010100 \cdots 00100110$};
    \node[left = of xors] (xors-label) {Xored Chunks};

    \node[arity = 2, below = 0.5em of xors] (masks) {
      $00011111 \cdots 00011111$
      \arg{1} \dots
      \arg{2} $00011111 \cdots 00011111$};
    \node[left = of masks] (masks-label) {Mask};

    \node[
      single arrow,
      anchor = west,
      yshift = -1em,
      rotate = 270,
      draw = green,
      fill = green,
      fill opacity = 0.75,
      minimum width = 10pt,
      single arrow head extend=3pt,
      minimum height=10mm
    ] (arrow-and) at (masks.south) {};
    \node[right = 0.25em of arrow-and.north] (and-label) {\scalebox{2}{$\otimes$}};

    \node[arity = 2, below = 1em of arrow-and.east] (mutations) {
      $00000000 \cdots 00010000$
      \arg{1} \dots
      \arg{2} $00010100 \cdots 00000110$};
    \node[left = of mutations] (mutations-label) {Mutation Edges};

    \node[
      single arrow,
      anchor = west,
      yshift = -1em,
      rotate = 270,
      draw = green,
      fill = green,
      fill opacity = 0.75,
      minimum width = 10pt,
      single arrow head extend=3pt,
      minimum height=10mm
    ] (arrow-tz) at (mutations.south) {};
    \node[right = 0.25em of arrow-tz.north] (tz-label) {\scalebox{1.5}{\code{trailing\_zeros}}};

    \node[below = 1em of arrow-tz.east] (result) {Number of mutations edges};
  \end{tikzpicture}
  \caption{Schema of the MMN algorithm on an AVX-512 compliant architecture. After the ancestral edges for the interval of interest $\omega$ have been listed, the haplotypes of their parent and child vertices are divided into chunks of 8 markers, which are stored in two separate arrays: the first chunk in the ``Parents'' array corresponds to the parent of the first edge, the second chunk to the parent of the second edge, and so on. The ``Children'' array is constructed similarly. Each matching pair of cells from both arrays is then xored in a single operation, resulting in the ``Xored Chunks'' array. The bitwise conjunction between each chunk and the mask $m_{\omega}$ is then computed in a similar fashion. Note that the ``Mask'' array exists for illustrative purposes only and does not need to be instantiated in practice. The number of mutation edges is obtained by repeated bitshift and application of \code{trailing\_zeros} on ``Mutation Edges''. The mutation edges can be obtained simultaneously via the indices of set bits.}
\end{figure*}

\begin{algorithm}
  \DontPrintSemicolon

  \SetKwData{Midx}{markeridx}
  \SetKwData{Cidx}{chunkidx}
  \SetKwData{Mask}{$m_{\omega}$}
  \SetKwData{Int}{$\omega$}
  \SetKwArray{Es}{edges}
  \SetKwData{E}{e}
  \SetKwArray{Css}{chunks\_src}
  \SetKwArray{Cds}{chunks\_dst}
  \SetKwArray{Mes}{mutationedges}
  \SetKwData{I}{i}
  \SetKwData{J}{j}
  \SetKwData{K}{k}
  \SetKwFunction{CTZ}{trailing\_zeros}
  \SetKwData{Pos}{pos}
  \SetKwFunction{ItP}{idx\_to\_pos}
  \SetKwFunction{Ai}{ancestral\_interval}
  \SetKwFunction{Push}{push!}
  \SetKwData{Acc}{acc}
  \SetKwData{MidxChunk}{mutationidx\_chunk}

  Set $\Midx \leftarrow 1$ \;

  \While{\Midx < \# markers} {
    Empty \Es, \Css, \Cds and \Mes \;

    \BlankLine
    Set \Cidx to the index of the chunk containing marker \Midx \;
    Compute $\Int$, the ancestral interval of unmasked markers in the chunk \;
    Compute \Mask to mask irrelevant markers to 0 \;

    \BlankLine
    Fill \Es, \Css and \Cds with edges and adjcent chunks by traversing the
      marginal graph associated with \Int bottom-up, stopping early when a chunk
      masked with \Mask is equal to zero. \;\label{line:mmn-store}

    \BlankLine
    \For{$\K \in 1, \ldots, |\Es|$} {
      \tcp{Lines 2 and 3 of \cref{alg:mmn-edge} (in-place)}
      Set $\Css{\K} \leftarrow \Css{\K} \oplus \Cds{\K}$ \;
      Set $\Css{\K} \leftarrow \Css{\K} \otimes \Mask$ \;
    }

    \BlankLine
    \For{$\I \in 1, \ldots, |\Es|$} {
      Set $\Acc \leftarrow 0$ \;

      \While{true} {
        Set $\J \leftarrow \CTZ{\Css{\I}}$ \tcp{Hardware operation}\label{line:mmn-ctz}
        \If{j > 8} {
          break \;
        }
        Set $\Pos \leftarrow \ItP{8(\Cidx - 1) + \Acc + \J}$ \;\label{line:mmn-idxtopos}
        \If{$\Pos \in \Ai{\E}$} {
          \Push{\Mes{\Acc + \J}, \E}
        }
        Update $\Css{\I} \leftarrow \Css{\I} \gg j$ \;
        Update $\Acc \leftarrow \Acc \gg j$ \;
      }
    }

    \BlankLine
    Set \MidxChunk to the index of the first entry of \Mes with cardinality
      strictly greater than one. If there is no such entry,
      $\MidxChunk \leftarrow 0$. \;
    \If{$\MidxChunk > 0$} {
      \tcp{ARG is not consistent; return index of the next inconsistent
        marker \& associated mutation edges}
      Return $8(\Cidx - 1) + \MidxChunk$, \Mes{\MidxChunk}
    }

    \BlankLine
    Set $\Midx \leftarrow 8 \Cidx + 1$ \;
  }
  \BlankLine
  \tcp{ARG is consistent}
  Return 0, $\emptyset$ \;

  \caption{MMN}\label{alg:mmn}
\end{algorithm}

\subsubsection{Breakpoint}\label{breakpoint}
Once an inconsistent marker has been found, a sequence of RR events is generated
in a way that reduces the local number of mutations down to one. The RR
procedure itself is discussed in details in \cref{recombination-recoalescence}.
For now, we are concerned with the distribution of breakpoints.

Standard theory models the occurrence of recombination events as a Poisson
process, which leads to an elementary algorithm for sequential simulators
revolving around sampling exponential interarrival times. Things are different
when inferring ARGs. Assuming a strong mutation model such as the ISM greatly
restricts the number and position of breakpoints. Let $b$ and $m$ be the
positions of the last breakpoint and the next inconsistent marker, respectively.
Since, under the ISM, some events must be positioned in this interval, it
would not make sense for the distance between $b$ and the next breakpoint to
follow an exponential distribution. A standard result about PPPs states that
events are uniformly distributed in an interval conditional on the number of
events in that interval. From this point of view, the correct way of choosing
breakpoints in $[b, m)$ would be to first draw a Poisson-distributed number of
events and then position them uniformly in $[b, m)$. Unfortunately, generating
a sequence of events that would yield a tree consistent with the marker at
$m$ having a predetermined length is prohibitively difficult unless we are
ready to either engage in additional time-consuming computation or exclusively
locate recombination events on \emph{derived edges}, that is those edges whose
downstream vertex (the one closer to the leaves)  has the derived allele at
the focal marker. As described in \cref{recombination-recoalescence}, the
maximum number of mutations eliminated by a recombination event occurring
on a \emph{wild edge}, that is, an edge whose downstream incident vertex is
associated with the wild allele for the marker under consideration, is the
number of mutations in the marker's marginal tree minus one. In other words,
depending on the ARG's topology, a constrained RR event might remove anywhere
from a single mutation to all of them.

An additional complication is that while choosing breakpoints in $[b, m)$ leads
to a reduction in the number of mutations for the marker at $m$ assuming correct
locations for RR events, there is no guarantee that doing so will not create
mutation edges for markers to the left of $m$. Since we are inferring ARGs from
left to right, increasing the number of mutation edges for markers to the right
of $m$ is not a problem. It would generally be desirable for that number to go
down, but designing a scalable algorithm to achieve that goal would be challenging, since mutation edges associated with every marker to the right of $m$ would need to be computed for multiple breakpoints in $[b, m)$ and, for each breakpoint, multiple sequences of RR events.
In any case, additional mutation edges created for markers to the right will be
dealt with in subsequent iterations. Creating mutations for markers that have
already been processed is more problematic. While we could backtrack to deal
with those, this would negatively affect performance and necessitate generating
additional events. Instead, \pkg{Moonshine} imposes additional constraints
on breakpoints' distribution. Let $m'$ be the position of the first marker to
the left of $m$. An event positioned in $[m', m)$ cannot create mutation edges
for markers to the left since any associated edge is not ancestral for those
markers.

Restricting the support of breakpoints to the interval between the next
inconsistent marker and the one directly to its left is efficient but overly
prohibitive. In addition to reducing the support of breakpoints for no reason
other than ease of implementation, it diminishes our algorithm's ability to
generate type 2 recombination events. Algorithms based on a first-order Markov
approximation of the recombination process, such as SMC and SMC', take a rather
drastic approach by simply ignoring any event that might have happened anywhere
except on the current marginal tree. Other algorithms allow recoalescence
events to happen further along the sequences while still restricting the
position of recombination events. The main selling points of these algorithms
are their computational efficiency and relative simplicity. However, since
recombination events are precluded from happening in non-ancestral material,
they are inherently incapable of generating type 2 recombinations. In order to
simulate these, a sequential algorithm must be able to go back in space, so to
speak, while simulating the recombination process: some events must happen to
the left of the preceding one. This class includes algorithms such as those implemented by
\pkg{SC} and, by extension, \pkg{SC-sample}. Their strategy is to generate a
recombination event on either the coalescence branch of the previous RR event
or one located upstream. This process is repeated until recoalescence with the
current marginal tree occurs. \pkg{ARGinfer} is another algorithm able to produce
type 2 recombination. Being an MCMC-based, it does so via two proposals:
``adding a new recombination to a lineage'' and ``resampling the breakpoint of a
recombination event''~\citep{Mahmoudi2022}.

These algorithms have different approaches to type 2 events. One
approach, used by \pkg{SC}, aims to stay closer to the coalescent process
by generating recombinations according to a complex distribution. The other,
used by \pkg{ARGinfer}, uses a less sophisticated distribution and relies on
the properties of its underlying MCMC sampler for statistical correctness.
It is apparently able to propose a move faster than \pkg{SC} can generate a
recombination event. However, many of those moves may be rejected before one is
finally accepted, while every event generated by \pkg{SC} is actually integrated
into the ancestry. Both approaches have their merits, and it would be risky to
make a unilateral statement as to which is the best.

Our package shares many characteristics of \pkg{SC} and \pkg{ARGinfer}, but its
approach to type 2 events is different. Since we are processing haplotypes from
left to right and only consider events resulting in a reduction of the marginal
number of mutation edges, breakpoints associated with type 2 events have to
be positioned to the left of not only $m$ but of $m'$ as well. This begs the
question: what are the conditions under which a recombination event positioned
at $b \leq b' < m'$ does not increase the number of mutation edges in $[b, m)$?
Let $e_r$, $e_c$ be the recombination and recoalescence edges respectively,
$v_r$, $v_c$ the associated child vertices, and $h_{v_r}^k$, $h_{v_c}^k$ the
status of the $k$\textsuperscript{th} marker of the associated haplotypes.
An RR event positioned at breakpoint $b'$ will not result in the creation of
additional mutation edges in $[b, m)$ as long as for every marker $k$ in $[b',
m)$, one of the following conditions is met:
\begin{enumerate}
  \item $h_{v_r}^k = h_{v_c}^k$;
  \item $e_r$ is not ancestral for marker $k$;\label{it:rec-type-2}
  \item $e_c$ is not ancestral for marker $k$ and the status of the first ancestral
    vertex upstream $v_c$ is $h_{v_r}^k$.
\end{enumerate}
Among these conditions, \cref{it:rec-type-2} allows for type 2 events. Since
we are working from left to right, support for those breakpoints comes with
the potential for type 4 events as well. Consequently, they must be
explicitly discarded.

Assuming the lower limit of the support for the $i$\textsuperscript{th}
breakpoint $b_i'$ is known, we need to decide on a distribution for the
breakpoint itself. Since the support is bounded, the two most natural choices
are the truncated exponential and uniform distributions. The main advantage
of the former is its resemblance to the untruncated exponential distribution,
which is the mathematically correct choice assuming a PPP and an unknown number
of recombination events in $[b, m)$. For minimum departure from this model, the
$i$\textsuperscript{th} breakpoint $b_i$ would be supported on $[\max \{b_i',
b_{i - 1} \}, m)$ where $b_0 = 0$. In addition to precluding type 2 events,
this distribution has the drawback of being difficult to deal with from a
numerical standpoint. Each draw reduces the support of the next, making standard
approaches to random generation such as rejection sampling and inverse transform
less efficient and/or more prone to instability due to divisions and the use
of transcendental functions. Consequently, \pkg{Moonshine} uses the simpler
uniform distribution. Each breakpoint is supported on the full interval $[b_i',
m)$ which avoids numerical instabilities and allows for type 2 events. This is
akin to trying to stay as close as possible to the PPP distribution assuming
(erroneously) a known number of events.

\subsubsection{Recombination and Recoalescence}\label{recombination-recoalescence}
When processing sequences from left to right, the procedure described in the
previous section allows to efficiently find the next site incompatible with
the current ARG, that is, the site with two or more mutation edges. It is
straightforward to keep track of these by storing them in a list, for instance.
The next step is then to add recombination events to the ARG in a way that
renders it consistent with the focal site.

\begin{figure*} 
  \centering
  \begin{subfigure}[t]{0.3\textwidth}
    \centering
    \includegraphics[width = \columnwidth]{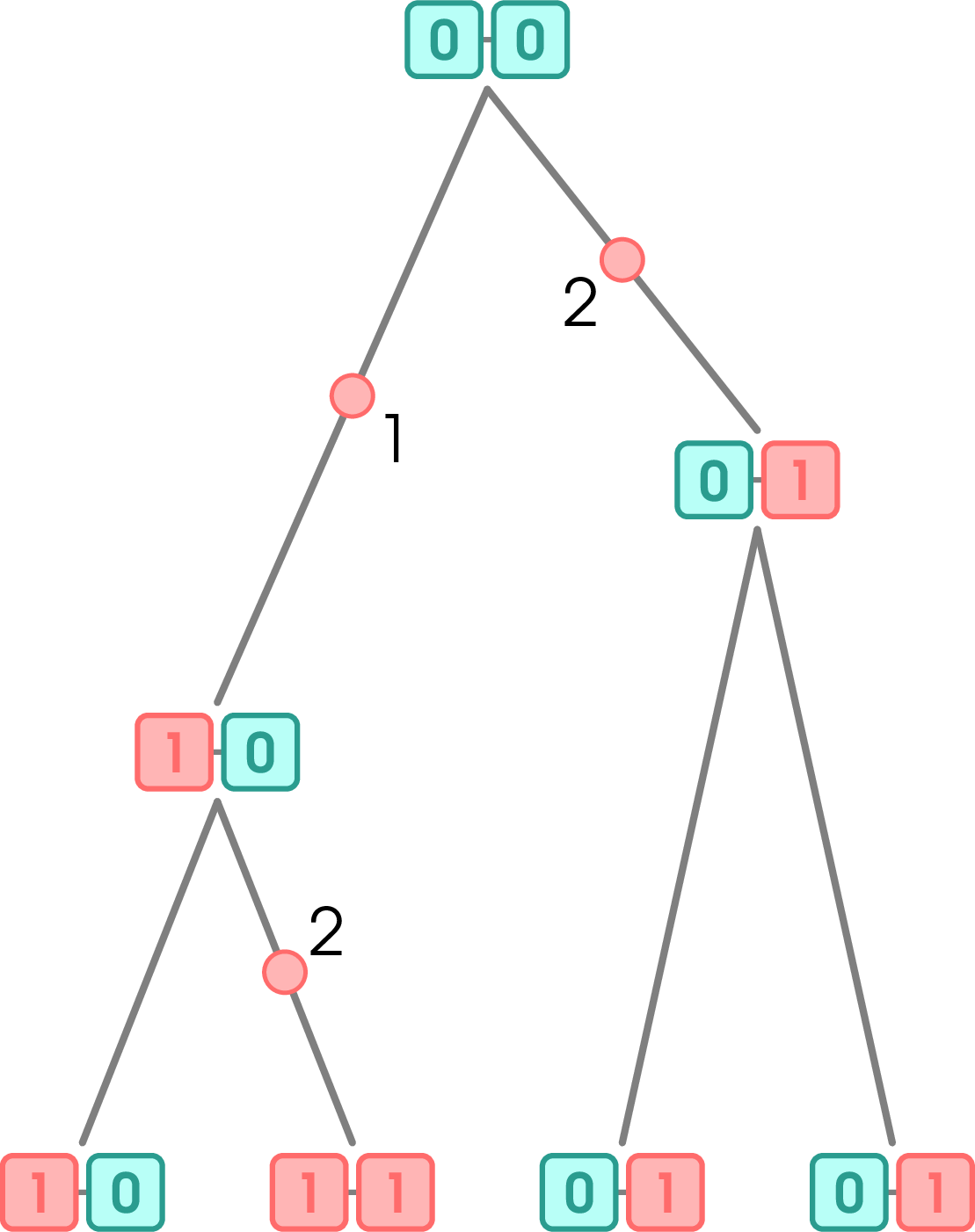}
    \caption{Inconsistent graph: marker 2 mutates twice.}\label{fig:derived-recombination-dd-1}
  \end{subfigure}
  \hfill
  \begin{subfigure}[t]{0.3\textwidth}
    \centering
    \includegraphics[width = \columnwidth]{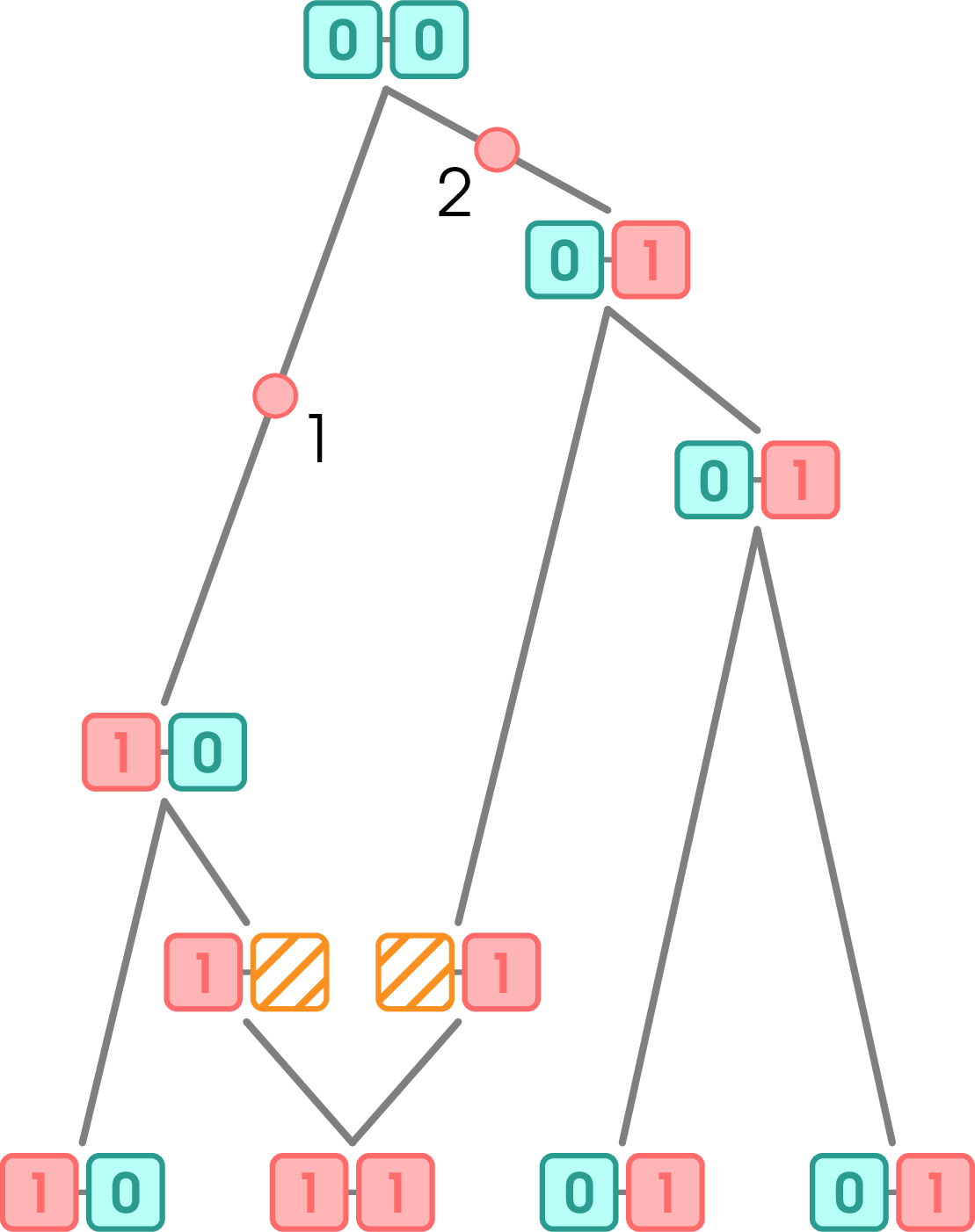}
    \caption{Recoalescence on a mutation edge. A single mutation is eliminated as a result.}\label{fig:derived-recombination-dd-2}
  \end{subfigure}
  \hfill
  \begin{subfigure}[t]{0.3\textwidth}
    \centering
    \includegraphics[width = \columnwidth]{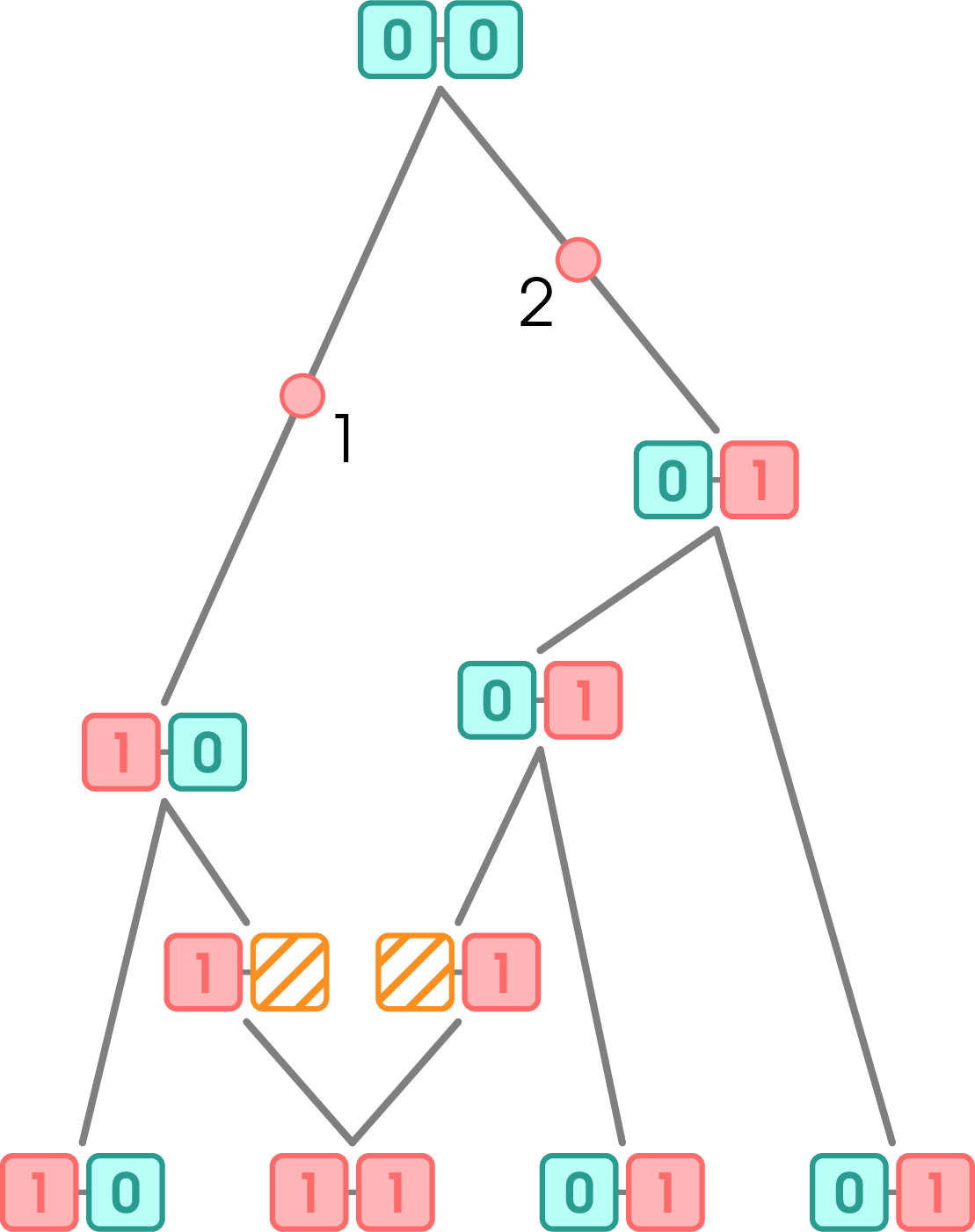}
    \caption{Recoalescence on a non-mutation derived edge. A mutation is eliminated even if the target of the recoalescence is not a mutation edge.}\label{fig:derived-recombination-dd-3}
  \end{subfigure}
  \caption{Derived recombination followed by a recoalescence event on a derived
  edge. Original graph is represented in \cref{fig:derived-recombination-dd-1}.
  \Crefrange{fig:derived-recombination-dd-2}{fig:derived-recombination-dd-3}
  are example of the two possible types of recoalescence events.}
\end{figure*}

Let $M$ be the set of mutation edges. The coalescence of two elements of $M$
results in a net diminution of the number of mutation edges by 1. The reason
is as follows: since the two coalescing edges have the same allele at the
focal site (namely the derived one), neither will be a mutation edge anymore
after coalescence. However, one edge upstream of the coalescence vertex will
ultimately coalesce with a wild edge and therefore become a new mutation edge,
resulting in the aforementioned net diminution. To conclude the argument,
note that it is not possible for a coalescence event to decrease the number of
mutation edges by more than 1. For that to be possible, one of the coalescing
edges would need to have a sibling that is itself a mutation edge. However,
in that case, the coalescing edge would not be a mutation edge. We call
recombination events located on a derived edge and the recoalescence event that
follows \emph{derived RR event}. A configuration resulting from such a pair of
events is given in \cref{fig:derived-recombination-dd-2}. The previous argument
actually entails an even stronger conclusion: there is no requirement for the
edge on which recoalescence occurs to be a mutation branch. As illustrated in
\cref{fig:derived-recombination-dd-3}, merely being derived is sufficient.

\begin{figure} 
  \centering
  \begin{subfigure}[t]{0.3\textwidth}
    \centering
    \includegraphics[width = \columnwidth]{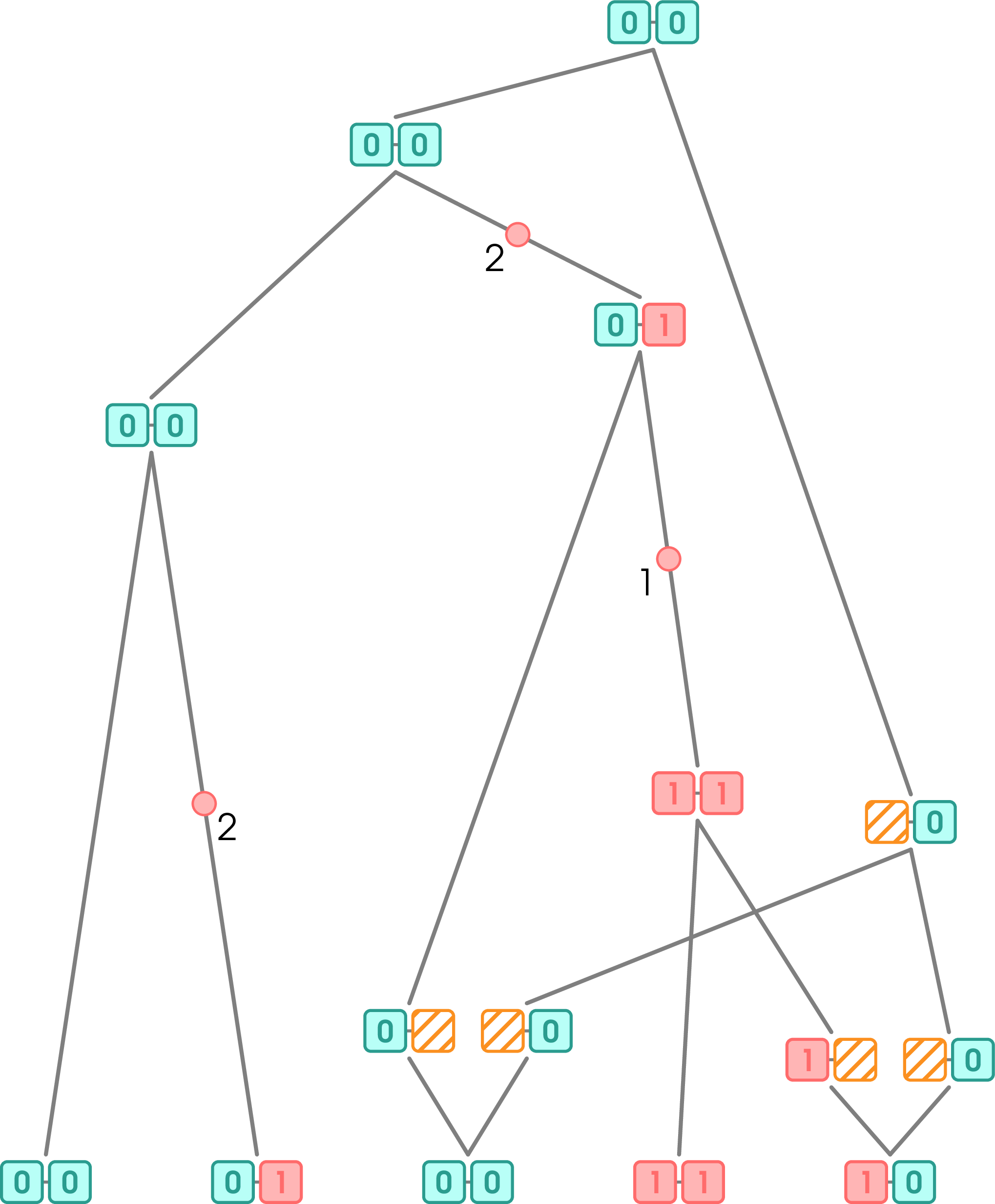}
    \caption{Inconsistent graph: marker 2 mutates twice.}
  \end{subfigure}
  \hfill
  \begin{subfigure}[t]{0.3\textwidth}
    \centering
    \includegraphics[width = \columnwidth]{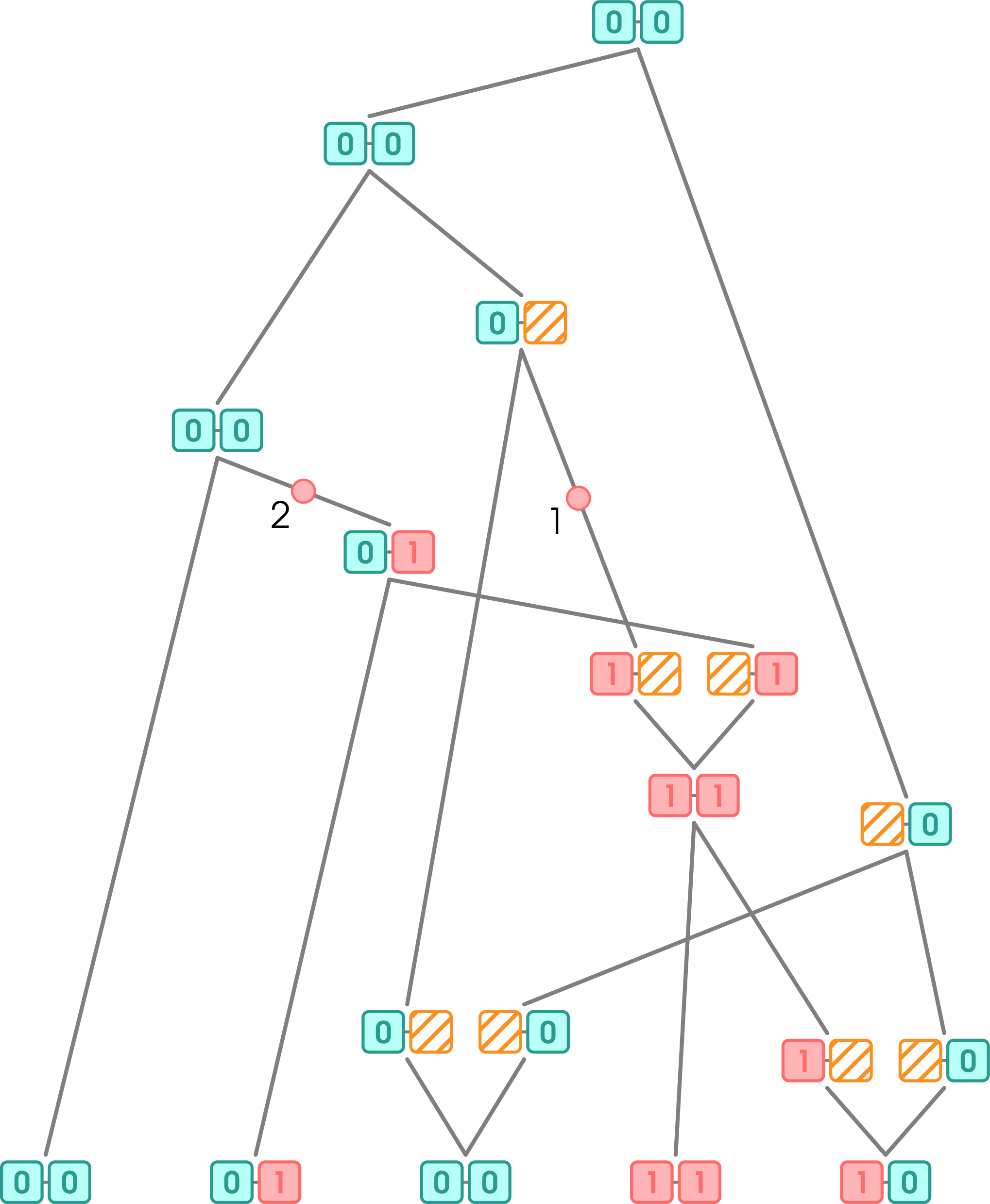}
    \caption{Recombination event on one of the child edge of a mutation edge,
    its sibling being non-ancestral for marker 2.}\label{fig:derived-recombination-dnd-2}
  \end{subfigure}
  \hfill
  \begin{subfigure}[t]{0.3\textwidth}
    \centering
    \includegraphics[width = \columnwidth]{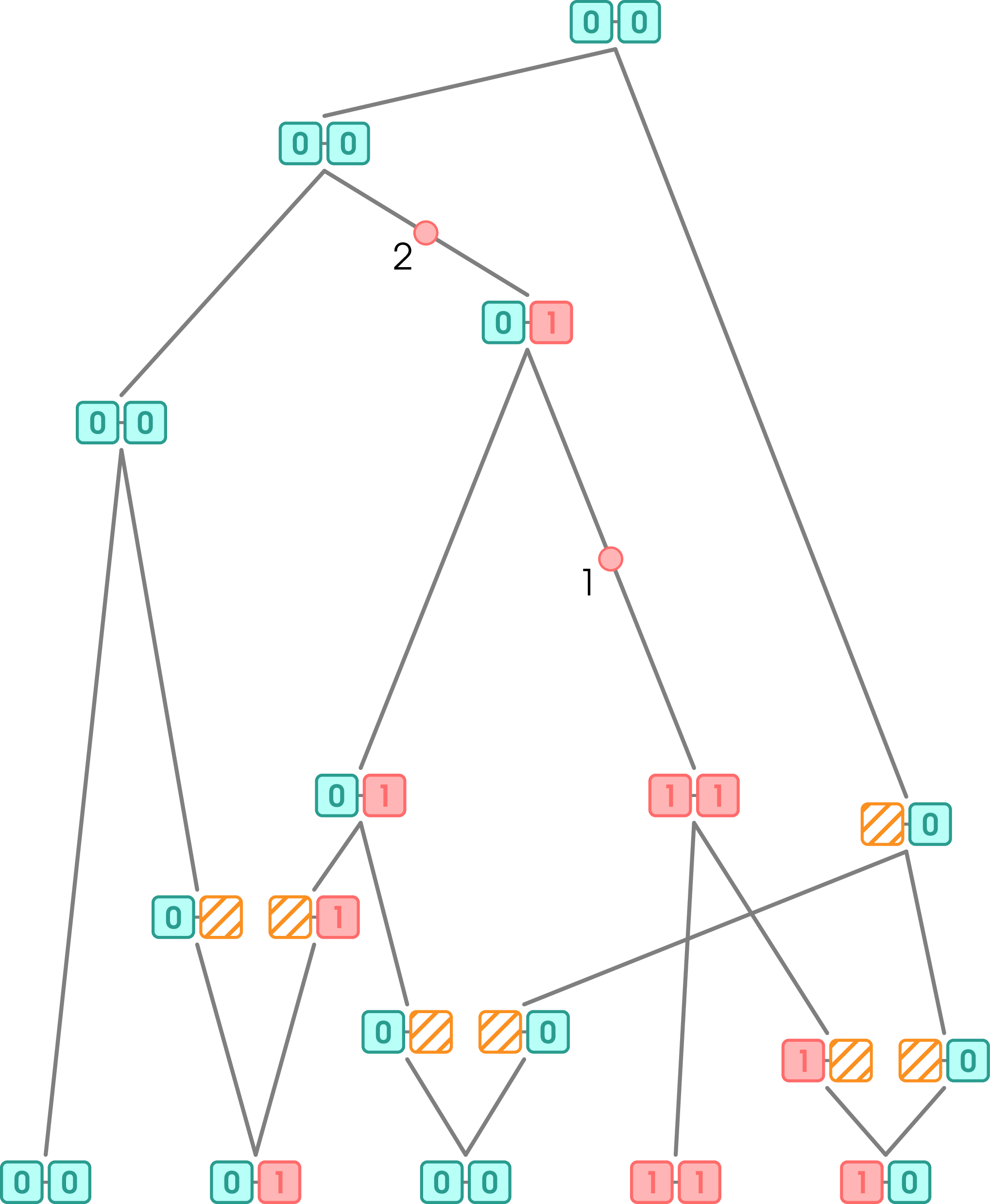}
    \caption{Recombination event on a mutation edge followed by recoalescence on
    a non-ancestral branch downstream of another mutation
    edge.}\label{fig:derived-recombination-dnd-3}
  \end{subfigure}
  \caption{Derived recombination event located on an edge downstream of a
  mutation edge. Reduction in the number of mutation is achieved since the
  recombination edge has no sibling marginally.}\label{fig:derived-recombination-dnd}
\end{figure}

The type of derived events we just described is said to be \emph{constrained}
because their locations and positions are supported on a subset of the ARG
branches and sequences, respectively. Ideally, this subset should encompass as
many branches as possible, making the distribution closer to the true (i.e.,
PPP) one. It is in fact possible to extend the set of candidate branches
without compromising the reduction of the number of mutations. For one thing,
recombination events do not have to be located on a mutation edge. Let $\edge u
v$ be such an edge with $v$ being the downstream vertex.
Since $v$ is derived, so is every vertex located downstream in the current
marginal tree. Let $E_v$ be the set of edges downstream of $v$ in that tree. The
number of mutations can be reduced by one by taking a subset of $E_v$ separating
$v$ from the ARG's leaves and generating an RR event on each edge, provided
that each recoalescence events is located on derived edges outside $E_v$. If we
want to avoid inflating the number of recombination events, we can impose the
additional restriction $|E_v| = 1$ and generate a recombination event on $\edge u
v$ if no such edge separator satisfies this constraint. This is easily achieved
in practice by looking for sequences
\begin{equation*}
  \edge u v, \edge v w_1, \edge{w_1}{w_2}, \ldots
\end{equation*}
where each edge is marginally without sibling. In particular, this is the
case if $v, w_1, w_2, \ldots$ are recombination vertices, although it is not a
necessary condition, as illustrated by \cref{fig:derived-recombination-dnd-2}.
Considering these edges as possible locations for recombination events allows
for a more even distribution, which, in addition to making the approximation
to the true distribution, contributes to the elimination of numerical errors
associated with a large number of such events relative to the total branch
length. Sequential algorithms come with something of a numerical curse: as
the ARG grows, so does the number of short branches resulting from events
 located close to a branch's incident vertices. Soon enough, numerical
instabilities arise when a short mutation edge is encountered. Fortunately, as
we have seen earlier, the remedy is simple to implement.

We go even further by applying a similar analysis to the location of
recoalescence events. As noted by~\cite{Wang2014}, these do not need to be
supported on derived edges. A recoalescence with a non-ancestral edge downstream
of a derived edge does not increase the number of mutations, a scenario
illustrated in \cref{fig:derived-recombination-dnd-3}. Since it is impossible
for a wild edge to lead to a derived one, the set of admissible locations for
recombination events can be described in rather simple terms: coalescence of
a derived edge with a mutation edge or any of its downstream edges does not
increase the number of mutations.

\begin{figure*} 
  \centering
  \begin{subfigure}[t]{0.45\textwidth}
    \centering
    \includegraphics[width = \columnwidth]{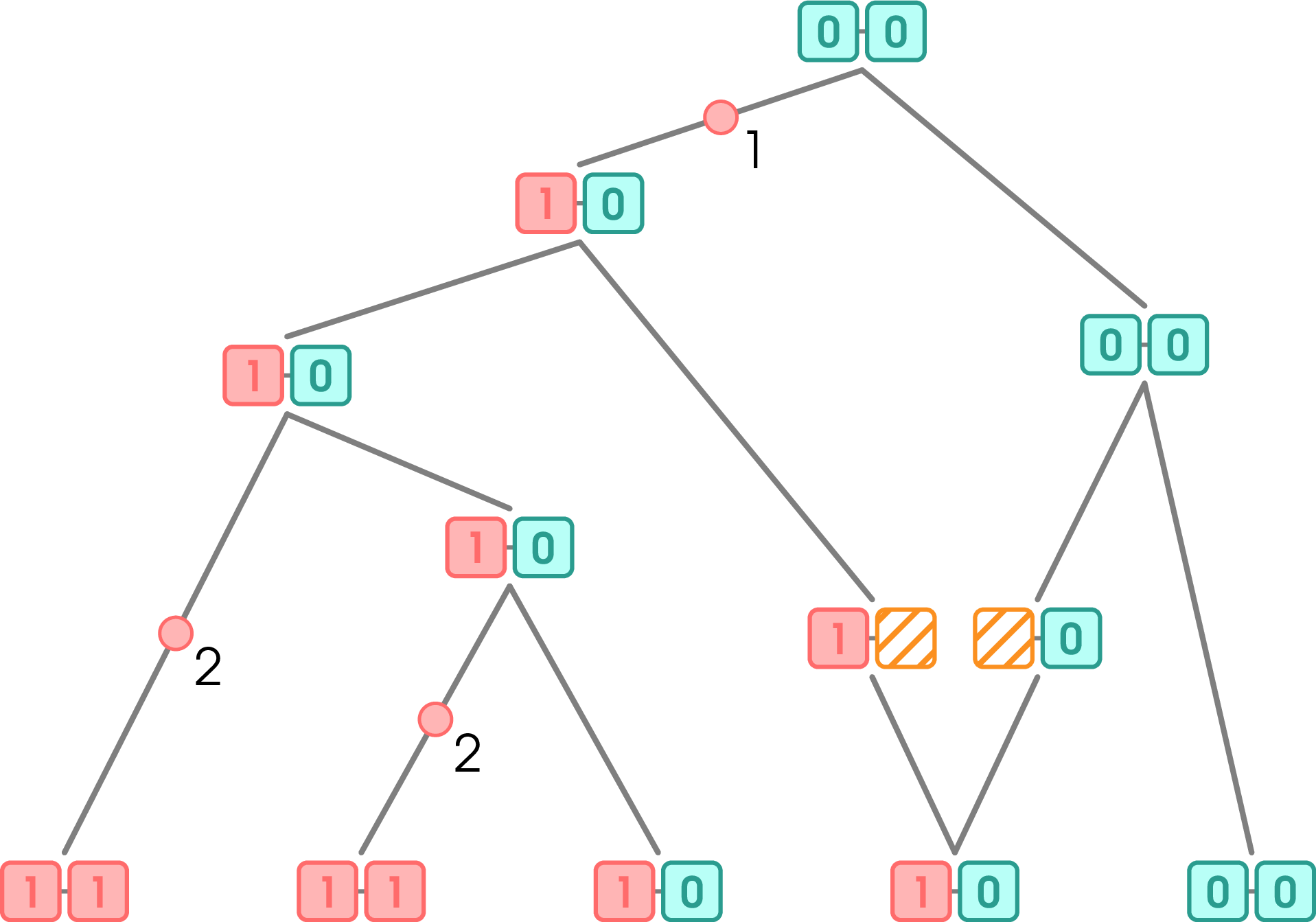}
    \caption{Inconsistent graph: marker 2 mutates twice.}
  \end{subfigure}
  \hfill
  \begin{subfigure}[t]{0.45\textwidth}
    \centering
    \includegraphics[width = \columnwidth]{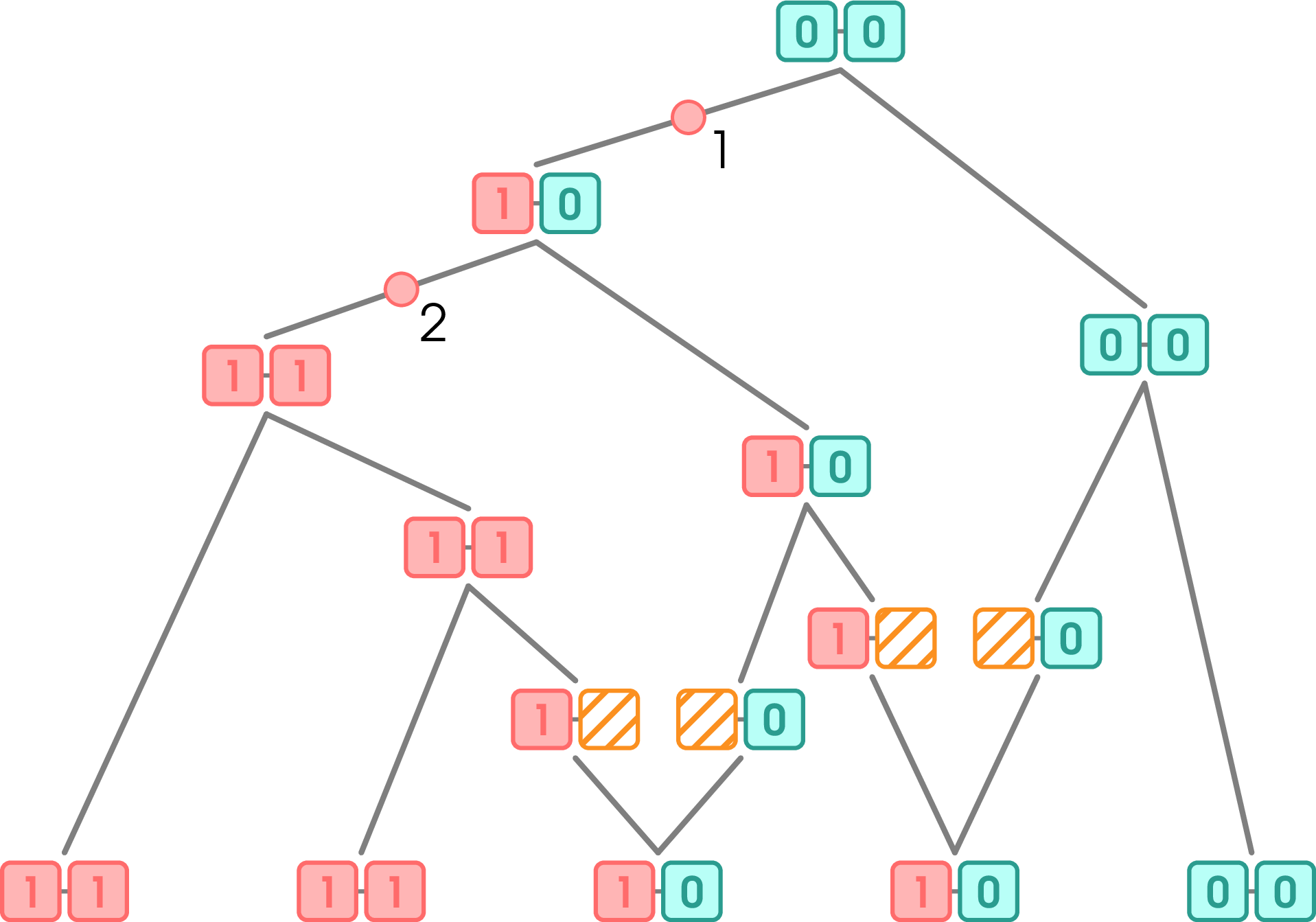}
    \caption{Wild recoalescence event (center vertex) with a non-ancestral edge leading to a wild branch.}
  \end{subfigure}
  \caption{Wild RR event leading to a reduction in the total number
  of mutations with recoalescence on non-ancestral
  edge.}\label{fig:wild-recombination-wn}
\end{figure*}

It should be clear by now that we can make any position consistent by generating
at most $|M| - 1$ derived recombination events. This is good, but we can
actually do better, at least some of the time. We refer to recombination events
located on a wild edge and the subsequent recoalescence event as \emph{wild RR
event}. Despite not occurring on a mutation edge, a wild recombination event
can effectively reduce the number of mutations when the following two conditions
are met:
\begin{enumerate}
  \item The recombination edge's brother is a mutation edge;
  \item Their uncle is a mutation edge.
\end{enumerate}
In such a configuration, the recombination event turns both derived vertices,
namely the sibling and uncle of the recombination vertex, into siblings with
respect to the current marginal tree, reducing the number of mutations by one.
As with the derived case, the wild recombination event does not need to occur
on the sibling of the mutation edge; the number of mutations will be reduced as
long as it is located downstream somewhere along a ``chain'' of edges without
marginal siblings. Just like in the derived case, coalescence can occur with
non-ancestral edges as long as they lead to a wild branch; this is illustrated
in \cref{fig:wild-recombination-wn}. Failing to meet this condition would result
in the creation of a new mutation event.

\begin{figure*} 
  \centering
  \begin{subfigure}[t]{0.4\textwidth}
    \centering
    \includegraphics[height = 1.5in]{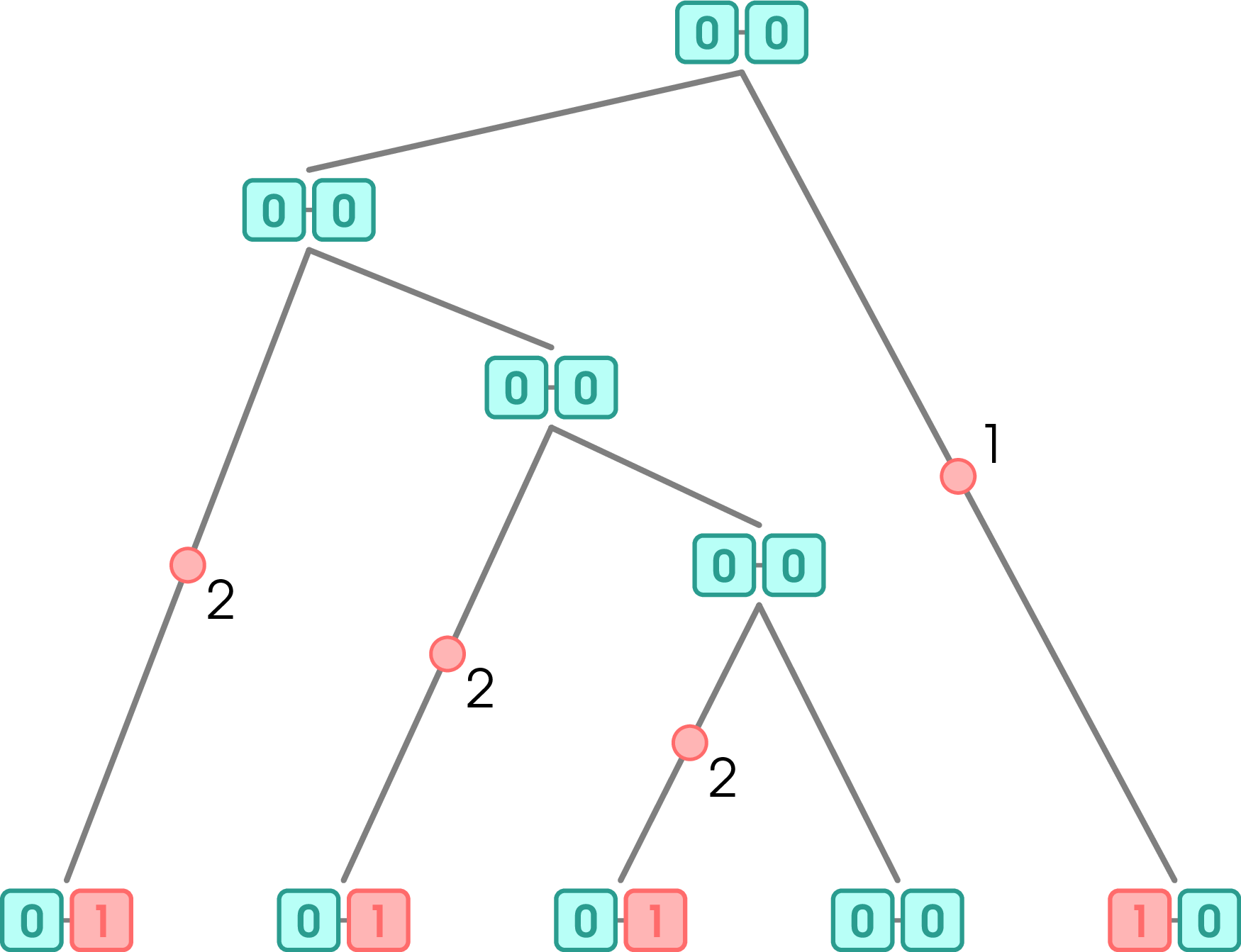}
    \caption{The second marker mutates three times on this genealogy. All these mutations result from the coalescence of the third and fourth leaves.}\label{fig:wild-recombination-chain-1}
  \end{subfigure}
  \hfill
  \begin{subfigure}[t]{0.4\textwidth}
    \centering
    \includegraphics[height = 1.5in]{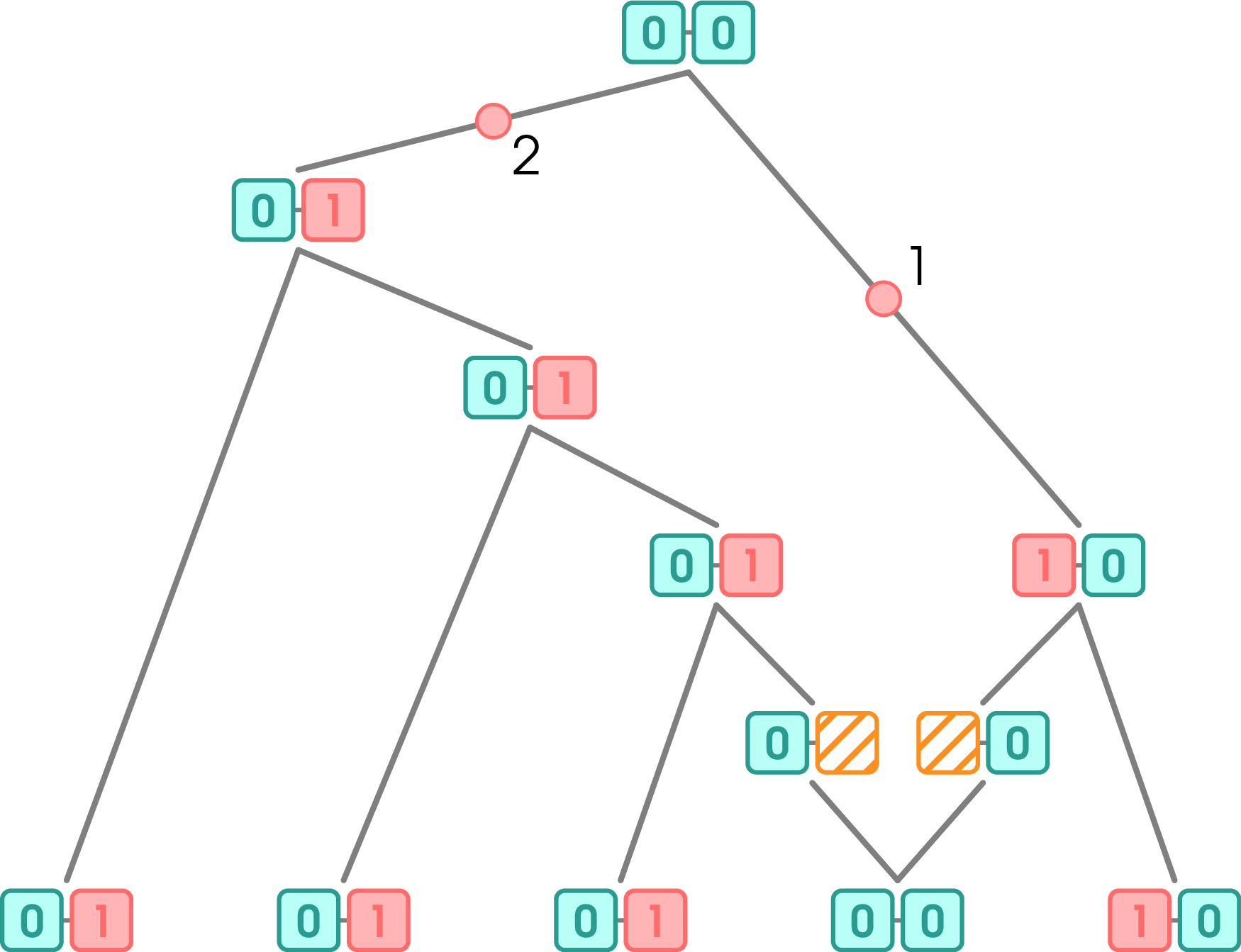}
    \caption{A recombination event prior to the coalescence of leaves 3 and 4 eliminates the mutation event on the branch incident to the third leaf. Consequently, mutation events directly upstream of leaves 1 and 2 are eliminated as well.}
  \end{subfigure}
  \caption{A single recombination event leads to the elimination of multiple
  mutations. This is only possible when the recombination event occurs on a wild branch adjacent to a series of ``derived cousins'', a configuration illustrated in \cref{fig:wild-recombination-chain-1}.}\label{fig:wild-recombination-chain}
\end{figure*}

When it comes to the number of mutations eliminated, wild RR events clearly
have the upper hand. As we just discussed, a wild recombination event flips the
allelic state of the mutation edge's parental vertex to derived. Consequently,
a reduction in the number of mutations can propagate upward as long as the
uncles encountered along the way in the current marginal tree are also mutation
edges. In fact, given an appropriate topology, a single mutation event can
reduce an arbitrary number of mutations to a single one. A scenario where a
single coalescence/recombination event leads to the elimination of two mutations
is illustrated in \cref{fig:wild-recombination-chain}. On the contrary, as
discussed before, the number of mutations eliminated by a derived RR event is
limited to one. This is because the mutation is eliminated by the recoalescence
rather than the recombination event. The state of the marker of the parental
vertex of the mutation edge on which the recombination event occurred remains
unchanged, inhibiting the propagation phenomenon observed in the wild case.

That being said, although \pkg{Moonshine} has the capability of inferring
ARGs by generating mutation reducing events exclusively, it does not actively
seek to maximize mutation reduction locally. It does not, for instance, give
priority to wild events over derived ones. As our main objective is to remain
close to the CWR, only branch length is taken into consideration when choosing
the location of a recombination event. Each edge on which a recombination
event would result in a reduction in the number of mutations for the current
marker has a probability of being chosen proportional to the difference in
latitude of its incident vertices, a strategy described earlier as
``constrained''. For reasons discussed earlier, the event's location on the
branch is not selected uniformly but rather assumed to be distributed as the same
location-scale family associated with the Beta distribution as the one used for
the unconstrained case.

Determining the location of a derived recoalescence event is fairly
straightforward. According to standard theory \citep{Wiuf1999}, coalescence
occurs at a unit rate with each admissible edge. We simply need to list all
available locations and pick one uniformly at random. The set of possible
recoalescence edges $E_R$ is readily established and does not require graph
traversal. The minimum latitude is either that of the recombination event or the
smallest latitude among the destination vertices of $E_R$, whichever is greater.
Each edge in $E_R$ is associated with a probability proportional to its length
minus any section outside of admissible latitudes. Once the recoalescence branch
is determined, a location is selected as usual on its admissible portion.

Simulating the location of a wild recoalescence event is more involved due
to the semi-infinite nature of its support. Its latitude is distributed as
the time of the first event of a non-homogeneous Poisson process. A common
approach in one-dimensional scenarios such as ours is the \emph{time-scale
transformation} method, which is a form of inverse transform sampling and
requires computation of the inverse of the integrated intensity function, also
known as the \emph{cumulative intensity function}. The main issue stems from
our decision not to track the number of live edges by latitude. This improves
general performance and reduces memory usage, but makes evaluation of the
intensity function extremely time-consuming, as it requires graph traversal. We
tackle this issue using numerical integration. Although somewhat variable, the
intensity function is piecewise constant and therefore a very good candidate for
quadrature. We use a logarithmic grid of latitudes to account for the generally
decreasing complexity of the ARG topology as height increases. The intensity
function is evaluated at quadrature nodes in a single partial traversal of the
ARG, reducing overhead to the minimum. By default, our algorithm uses a grid of
compile-time constant size 25, but this number can be tuned to balance precision
with performance using the preference mechanism.

\subsubsection{Multiple Crossing Over}
\begin{figure*} 
  \centering
  \begin{subfigure}[t]{0.4\textwidth}
    \centering
    \begin{tikzpicture}
      \node (figure) {
        \includegraphics[width = \columnwidth]{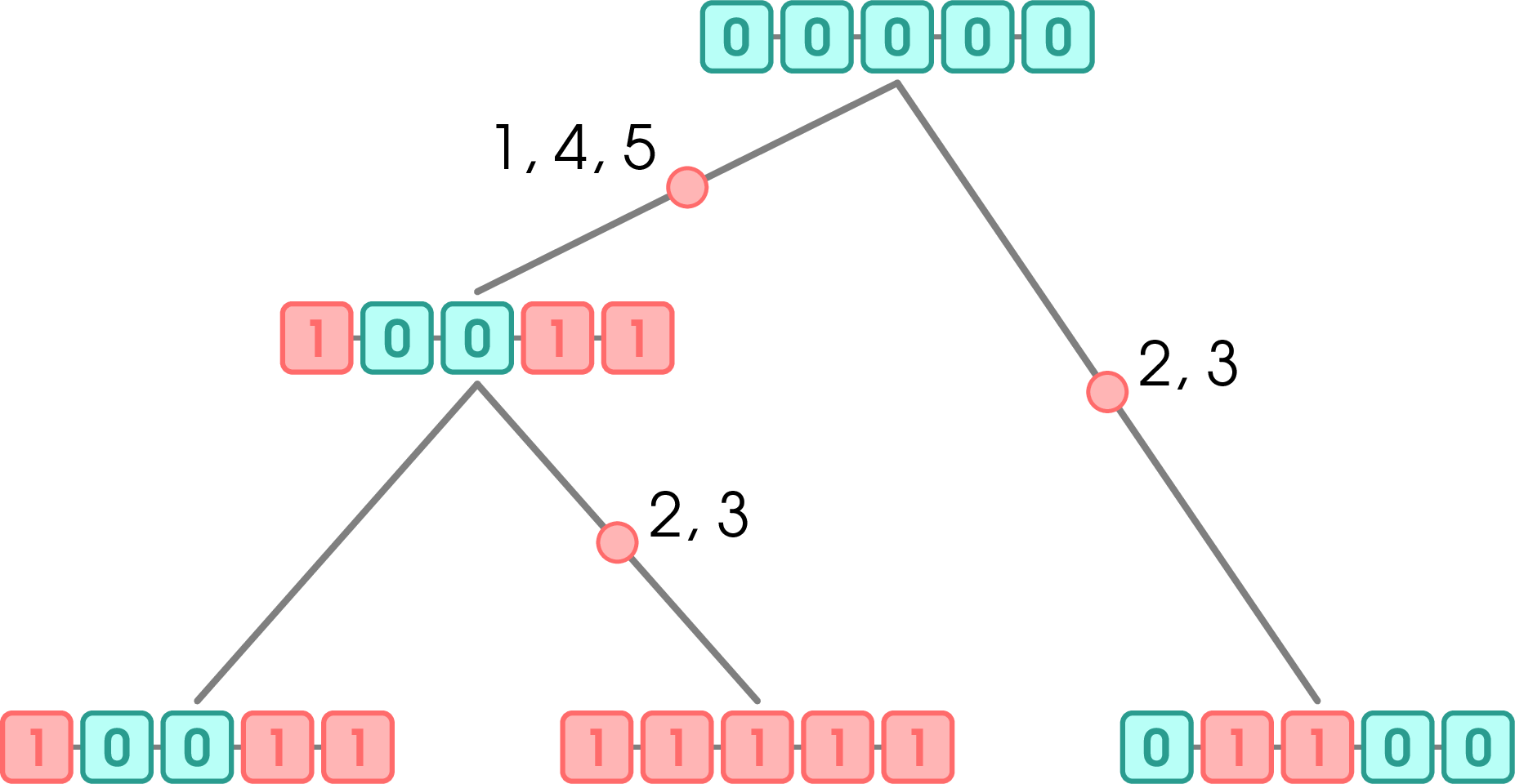}
      };
      \node[isosceles triangle,
            draw = recarrow,
            below = of figure,
            inner sep = 0pt,
            rotate = 90,
            fill = recarrow,
            opacity = 0,
            minimum size = 2mm] at (0, -1) {};
    \end{tikzpicture}
    \caption{Before recombination, markers 2 and 3 mutate twice.}
  \end{subfigure}
  \hspace{2em}
  \begin{subfigure}[t]{0.4\textwidth}
    \centering
    \begin{tikzpicture}
      \node (figure) {
        \includegraphics[width = \columnwidth]{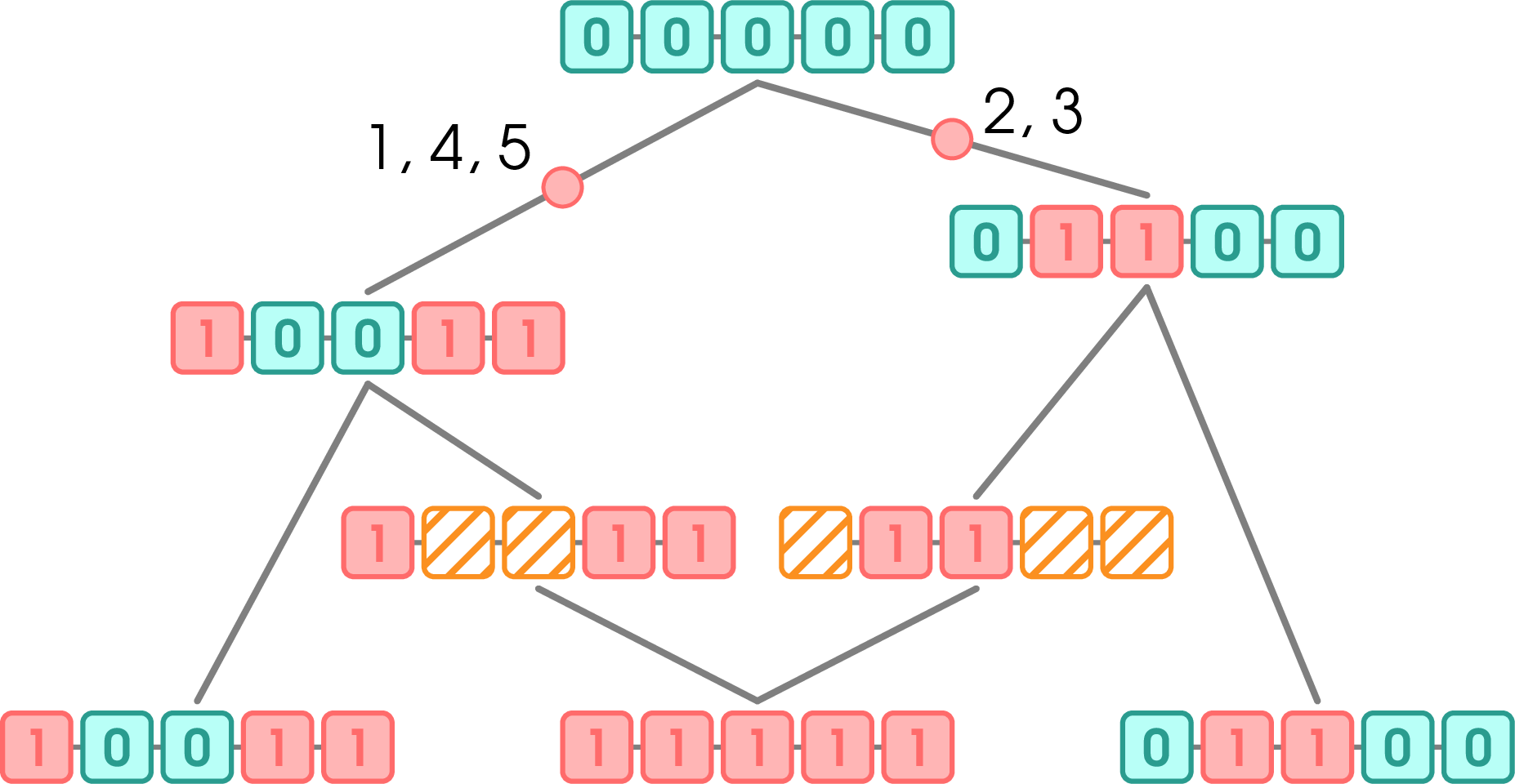}
      };
      \node[isosceles triangle,
            draw = recarrow,
            below = of figure,
            inner sep = 0pt,
            rotate = 90,
            fill = recarrow,
            minimum size = 2mm] at (-0.75, -1) {};
      \node[isosceles triangle,
            draw = recarrow,
            below = of figure,
            inner sep = 0pt,
            rotate = 90,
            fill = recarrow,
            minimum size = 2mm] at (0.2, -1) {};
    \end{tikzpicture}
    \caption{Consistency is achieved after two crossovers between markers 1 and 2, and 3 and 4.}
  \end{subfigure}
  \caption{Multiple crossover events. Breakpoints are indicated by pink arrows.}\label{fig:mco}
\end{figure*}

In addition to RR events, \pkg{Moonshine} can generate multiple crossing over
 events constrained to reduce the number of mutations. These events arise
in the following scenario: assume that the branch selected to undergo recombination
is the right parental edge of a recombination vertex. Assume
further that a recoalescence with the sibling of the other parental edge's
branch would reduce the number of mutations. When both of these conditions are
met, RR events are unnecessary; as illustrated by \cref{fig:mco}, the same effect can be achieved by modifying
the ancestral intervals of the parental edges of the recombination vertex. For
this to work, it is necessary to track the partition induced by recombination
events. In an \code{ARG}, every recombination vertex is associated with two sets
of intervals, one for each parental edge. We call these sets \emph{recombination
masks} and denote the mask associated with recombination $k$ by $m_k = \{ m_k^l,
m_k^r \}$. Initially, before any MCO event involving a specific recombination
vertex, the mask is rather simple. Let $b$ be the position of the associated
recombination event.
\begin{align*}
  m_k^l &= [0, b)\\
  m_k^r &= [b, \infty)\,.
\end{align*}
A MCO event at position $b' > b$ transforms those sets as follows:
\begin{align*}
  m_k^l &= [0, b) \cup [b', \infty)\\
  m_k^r &= [b, \infty) \cap [0, b') = [b, b')\,.
\end{align*}
A subsequent event at position $b'' > b'$ would yield
\begin{align*}
  m_k^l &= [0, b) \cup ([b', \infty) \cap [0, b'')) = [0, b) \cup [b', b'')\\
  m_k^r &= [b, b') \cup [b'', \infty)\,.
\end{align*}
In general, the correct mask can be computed by intersecting the rightmost
interval (with respect to the right endpoint) with $[0, b')$ and taking the
union with the other interval and $[b', \infty)$. Although explicit storage is
not necessary, coalescence vertices can be thought of as being associated with
the mask $[0, \infty)$. Recombination masks are used to compute the ancestral
intervals of parental edges in the following way: an edge's ancestral interval
is equal to the intersection of its recombination mask with the union of its
children's intervals.

Conditions under which a MCO event can occur are very restrictive. Consequently,
we expect detection to be underpowered. This might be improved in the future,
for example, by allowing the user to bias the procedure in favour of these events.

\subsection{ARG Update}
RR and MCO events both modify sequences and ancestral intervals associated with
vertices and edges upstream. The affected elements must be updated immediately
to ensure the soundness of the remainder of the procedure, which can be achieved
by traversing the ARG from the recombination and recoalescence edges toward the
root. Sequences and intervals updates are, however, computationally demanding.
Our first, rather naive implementation was a major bottleneck of the constrained
recombination algorithm. To make it as efficient as possible, we limit the
update procedure to the elements affected by the event. Recombination events
are mostly local, meaning they can only affect vertices and edges located
upstream. Consequently, any element not upstream of either the recombination
or recoalescence edge can be ignored when updating. In fact, an element must be
located upstream of a modified element to be modified itself. This means that
the number of updates can be further reduced by keeping track of the state of
every vertex and edge before they are updated and stopping when a match between
original and updated versions is detected. The algorithm terminates early if
both the sequence and the ancestral interval associated with an edge are left
unchanged.

The early termination strategy described above dramatically decreases the time
dedicated to ARG update. Indeed, coalescences with vertices left untouched
limit the spread of changes, often well below the root. It might be conceived
that the additional costs associated with storing information about elements
before updating them would outweigh the benefits of reducing the number of
updated elements. It turns out that a very significant number of recombination
events are very local in nature, to such a degree that we have yet to find
the point of diminishing return for reasonably large samples. That is not to
say, however, that the procedure cannot be improved further. We were able to
squeeze even more performance out of it through hashing. Instead of making a
copy of the current sequence and ancestral interval before update, we simply
compute a hash value for each of those, which we promptly hash together. The
procedure is terminated early if the hash of the updated sequence-ancestral
interval pair is equal to the original. Since hash functions are not injective,
there is a possibility that different pairs may have the same hash, an event
known as a \emph{collision}. Fortunately, this has not been a problem in
practice. We use a universal hash function from the 64-bit NH family \cite{Black1999} which have a pairwise collision probability of at most $2^{-64}$. To put any doubt to rest, the method \code{validate} can be applied
to the final product of the ARG inference process to ensure that it is exempt,
among other things, of the inconsistencies that would emerge from collisions. Future versions will allow users to select from multiple hash functions within the 64-bit NH family to mitigate the effects of a collision.
The complete procedure is presented in \cref{alg:arg-update}. Note that since
recombination vertices have a unique child, their associated sequence can be
a reference to their child's. Consequently, line \ref{line:arg-update-ref}
can be performed without additional allocation, dramatically cutting down
on memory usage. The early termination strategy is implemented by line
\ref{line:arg-update-early-term}.

\begin{algorithm}
  \DontPrintSemicolon

  \SetKwData{Stack}{stack}
  \SetKwData{svertex}{starting\_vertex}
  \SetKwData{H}{h}
  \SetKwData{E}{e}
  \SetKwFunction{Pop}{pop!}
  \SetKwFunction{Push}{push!}
  \SetKwFunction{Src}{src}
  \SetKwFunction{Parent}{parent}
  \SetKwFunction{Parents}{parents}
  \SetKwFunction{Haplotype}{haplotype}
  \SetKwFunction{Hash}{hash}
  \SetKwFunction{AI}{ancestral\_interval}

  Set $\Stack \leftarrow [\svertex]$ \;
  \While{\Stack is not empty} {
    Set $\V \leftarrow \Pop{\Stack}$ \;
    \uIf {\V is a recombination vertex} {
      Set the sequence of \V to that of its child \;\label{line:arg-update-ref}
      Set the ancestral interval of its parent edges to the intersection of its
        child's with the appropriate ancestral mask \;
      \Push{\Stack, \Parents{\V}}
    }
    \Else {
      Set $\H \leftarrow \Hash{\Haplotype{\V}}$ \;
      Set the sequence of \V to the conjunction of its children's haplotypes \;

      \BlankLine
      \If{\V is the root} {
        Continue \;
      }
      Set \E to the parent edge of \V \;
      Update $\H \leftarrow \Hash{\H, \AI{\E}}$ \;
      Set the ancestral interval of \E to the union of the ancestral intervals
        of \V's child edges \;

      \BlankLine
      \If {$\H \neq \Hash{\Hash{\Haplotype{\V}}, \AI{\E}}$} { \label{line:arg-update-early-term}
        \Push{\Stack, \Parent{\V}} \;
      }
    }
  }

  \caption{ARG Update}\label{alg:arg-update}
\end{algorithm}
\subsection{Technical Considerations}
\subsubsection{Markov Approximation}
\begin{figure}
  \centering
  \includegraphics[width = \columnwidth]{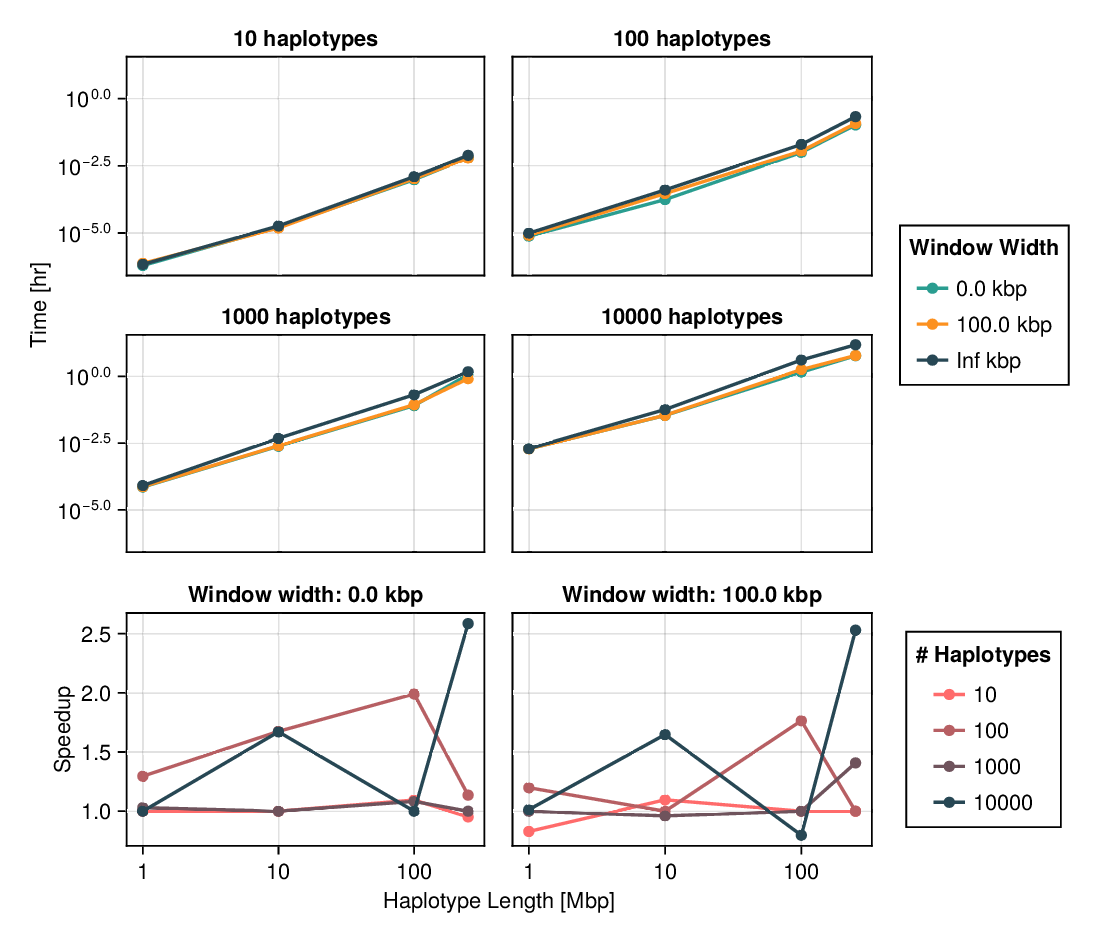}
  \caption{
    Time required to construct a single consistent ARG as a function of the number of haplotypes and haplotype length, including time spent inferring the initial tree. The speedup provided by two window widths with respect to an infinite window is depicted in the bottom two panels. Technical details are discussed in \cref{simulations}.
  }\label{fig:winwidth-benchmark-time}
\end{figure}

The procedure described in this section is exact in the following sense: at
any step, any edge of the ARG can undergo recombination or recoalescence
subject only to the mutation number reduction constraint. It is more
similar to the original Wiuf-Hein sequential algorithm than any of its Markovian
approximations, such as the SMC or SMC'. This makes ARGs more realistic,
but it comes with a performance penalty: as RR events are incorporated into
the graph, the number of edges that need to be considered for the next events
increases. Consequently, we should expect the number of operations performed
by our algorithm to grow as it progresses along sequences. One way to alleviate
this computational burden is to make the recombination process artificially
Markovian. The procedure we just presented allows for such an approximation,
although it was omitted from the description for simplicity. Both restricted
and unrestricted RR events may be executed inside a window moving across
the sequences, the width of which can be selected by the user. A width of 0
corresponds to a first-order Markovian approximation akin to the SMC, while an
infinite width means no approximation at all. In general, assuming distances in
base pairs (bp), specifying a width of $w$ for an RR event occuring at position
$p$ has the effect of excluding any edge $e$ such that
\begin{equation*}
  \ai(e) \cap [p - w, p + w] = \emptyset.
\end{equation*}
from the set of candidate recombination and recoalescence edges. We emphasize
that the window is \emph{centered} on $p$, as our algorithm is designed for
the general task of rendering ARGs consistent rather than building them from
a tree in a left-to-right sweep. Moreover, when generating the recoalescence
latitude of an unconstrained event, ignored edges are not taken into
account for the computation of the rate of the recombination latitude.
\Cref{fig:winwidth-benchmark-time} shows that the performance impact of
reducing the window's width is significant. For some combinations of sample
size and haplotype length, a window width of 0 can lead to over twice as
fast computation. Interestingly, the speedup is similar for a width of
100 kbp, suggesting that even relatively modest approximations can lead to
substantial reduction in computational resources. Irrespective of window size,
\cref{fig:winwidth-benchmark-time} suggests that computation time is exponential
in both the number of sampled haplotypes and markers.

\begin{figure}
  \centering
  \includegraphics[width = \columnwidth]{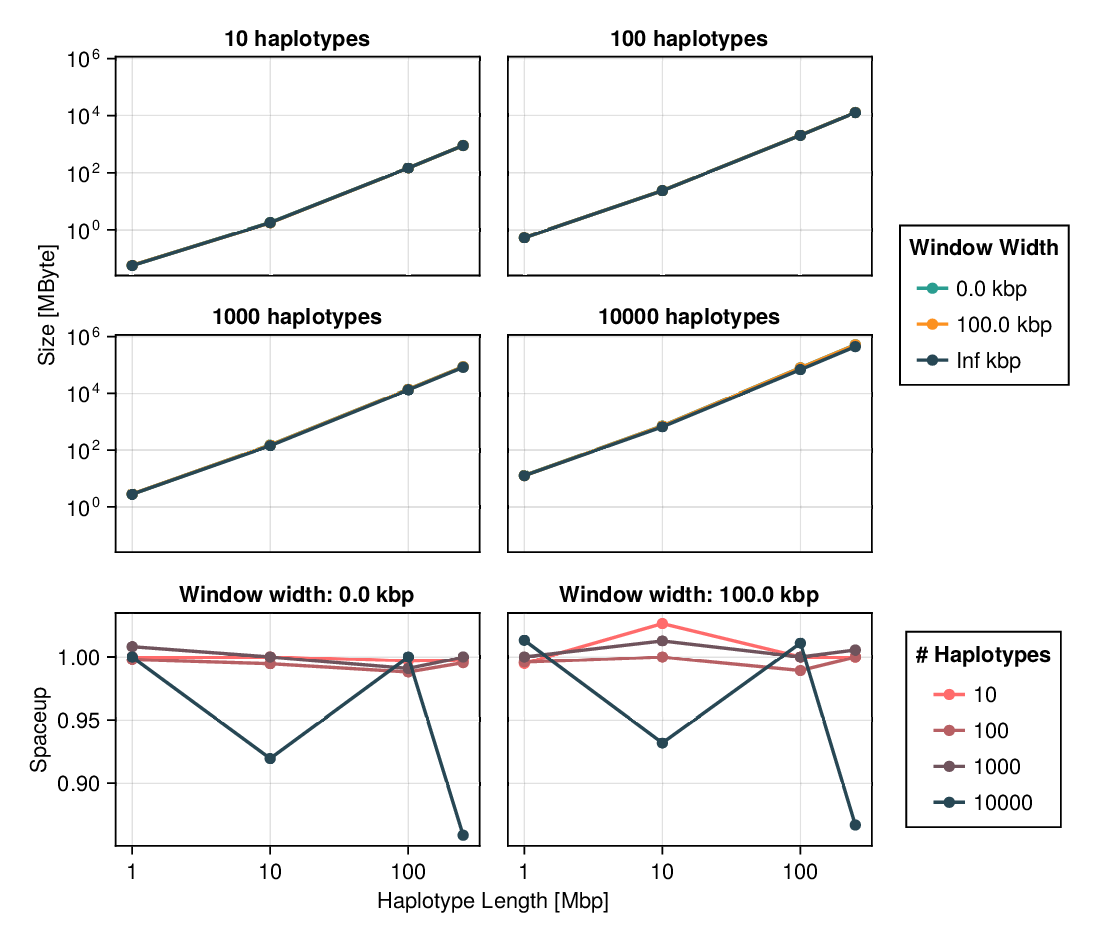}
  \caption{
     Size of the object containing the inferred ancestral recombination graph as a function of the number of haplotypes and haplotype length. The spaceup, plotted in the bottom two panels, is the ratio of the size for infinite window width versus that for the indicated width. Technical details are discussed in \cref{simulations}. We emphasize that the total amount of memory required by the procedure exceeds quantities reported here.
    }\label{fig:winwidth-benchmark-memory}
\end{figure}

\begin{figure}
  \centering
  \includegraphics[width = \columnwidth]{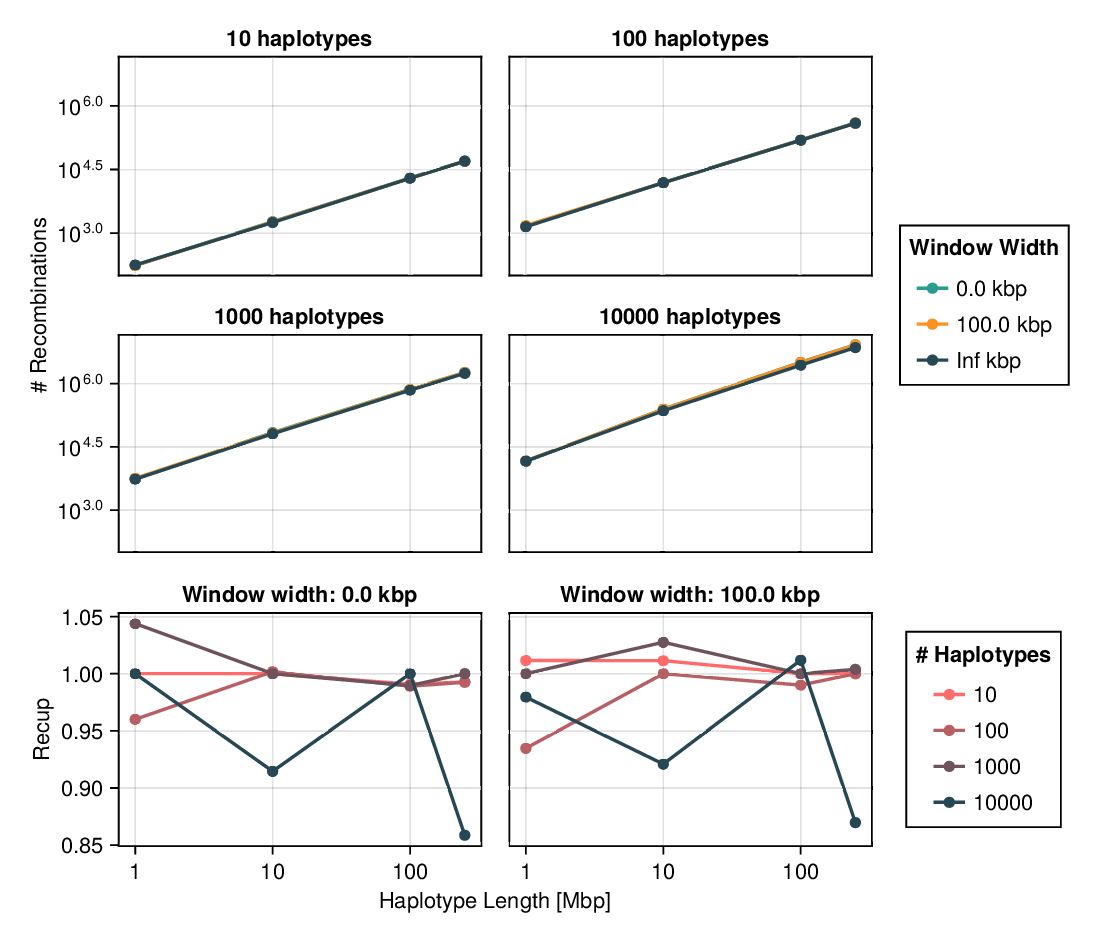}
  \caption{
    Number of recombination events generated by the ARG inference procedure as a function of the number of haplotypes and haplotype length. The recup, plotted in the bottom two panels, is the ratio of that number for an infinite window versus that for the indicated width. Technical details are discussed in \cref{simulations}.
  }\label{fig:winwidth-benchmark-recombinations}
\end{figure}

Similarly, \cref{fig:winwidth-benchmark-memory} suggests exponential
growth in memory usage. It appears, however, that reducing the
window size increases the size of the resulting ARG. This is largely
explained by the resulting increase in the number of recombination events. As shown in
\cref{fig:winwidth-benchmark-recombinations}, this number tends to increase
as the window width diminishes. Part of this phenomenon might be explained by
increased flexibility. Larger window sizes increase the number of candidate
edges for recombination and recoalescence events, allowing for constrained
coalescence of more similar haplotypes and, ultimately, more parsimonious
ancestries.

\subsubsection{Markers and positions}\label{markers-positions}
In addition to the ARG update procedure described at the beginning of this section, another major bottleneck in our method was, somewhat unexpectedly, a function called \code{pos\_to\_idx} that returns the index of a marker, given its position. It is the inverse of another function called \code{idx\_to\_pos}, which is designed solely to return the position of a given marker, indexed from the leftmost one. The term ``inverse'' in that context is used loosely, as multiple positions are associated with a given marker. For that reason, we define \code{pos\_to\_idx} formally as a pseudoinverse of \code{idx\_to\_pos}:
\begin{equation*}
  \postoidx(p) = \sup\{i \in 1, \ldots, s : \idxtopos(i) \leq p \}\,.
\end{equation*}
Since it is often necessary to mask sequences with respect to an ancestral interval, this function is used extensively within \pkg{Moonshine}'s recoalescence and ARG update procedures. The endpoints of any given interval rarely correspond to the position of a marker. This is compounded by the fact that what we refer to as ``ancestral interval'' is, in reality, the union of multiple disjoint intervals. Overall, a simple linear search is insufficient; a more efficient approach is called for. Binary search yields considerable gains, but we can do better. Memoization is a standard strategy for solving similar problems, but it is detrimental to our use case. The cost of computing \code{idx\_to\_pos} via binary search is small, even in comparison with a lookup in a hash table. Besides, positions are encoded as double-precision floating-point values, which limit their usefulness as keys of an associative structure. Perhaps it would be possible to use locality-sensitive hashing to circumvent this issue. We did not explore this avenue, though, because the approach currently implemented, although relatively simple, proved satisfactory. It is essentially a slightly optimized version of interpolation-sequential search~\citep{journals/ipl/GonnetR77}. We take advantage of the fact that the markers' position vector is not modified by ARG inference. Instead of directly computing the inverse of \code{idx\_to\_pos}, we do so on a first-order approximation obtained by least-squares fitting, which is straightforward. Even though computing such an approximation is time-consuming relative to an iteration of bisection search, it only has to be done once, making the overhead negligible. Pseudocode for \code{pos\_to\_idx} is given in \cref{alg:postoidx}.

\begin{algorithm}
  \DontPrintSemicolon

  \SetKwProg{Fn}{Function}{}{}

  \SetKwData{A}{a}
  \SetKwData{I}{i}
  \SetKwData{B}{b}
  \SetKwData{P}{p}
  \SetKwData{S}{s}
  \SetKwData{L}{l}
  \SetKwData{R}{r}
  \SetKwFunction{Idxtopos}{idx\_to\_pos}
  \SetKwFunction{Postoidx}{pos\_to\_idx}

  \tcp{Assume $\Idxtopos{\I} \approx \A \I + \B$}
  \BlankLine
  \Fn{\Postoidx{\P}}{
    Set $\I \leftarrow \frac{p - \B}{\A}$, $\L \leftarrow 1$, $\R \leftarrow \S$ \;
    Clamp \I in $[1, \S]$ \;
    Update \L or \R through a single binary search iteration \;
    Perform interpolation-sequential search with \I, \L and \R \;
  }

  \caption{pos\_to\_idx}\label{alg:postoidx}
\end{algorithm}

While \cref{alg:postoidx} is relatively simple, it takes advantage of three features of the positions vector: its static nature, its monotonicity, and the approximate regularity of the distance between markers. A more efficient procedure could likely be designed, but \cref{alg:postoidx} is at least efficient enough to eliminate the \code{pos\_to\_idx} bottleneck.

\subsubsection{Memory Allocation}
Considerable effort has been devoted to using memory as efficiently as
possible in the computationally demanding sections of \pkg{Moonshine}. In
addition to the obvious benefits in terms of reduced memory requirements,
we found dynamic allocation to be a major performance bottleneck in
itself. Although \proglang{Julia} is a high-level programming language,
fine-grained memory management is straightforward. It is entirely
possible to manage allocations directly. Part of the C standard library,
comprising functions such as \code{malloc}, \code{calloc} and \code{free},
is exposed to the user via the \pkg{Libc} module included in Julia's
standard library. Our approach, however, is slightly higher level: we make
extensive use of slab allocation~\citep{conf/usenix/Bonwick94} through
\pkg{Bumper.jl}~\citep{Protter2025}. This package allows us to handle raw
pointers when needed while efficiently managing the underlying memory pool
automatically. This allows for nearly effortless tight memory management
and dramatically reduces the number of allocations needed in core methods.
Performance-critical methods accept the \code{buffer} keyword argument through
which the caller can pass a \pkg{Bumper.jl} buffer. This is how the memory
is allocated by the \pkg{build!} method for inferring ARG for instance. The
same buffer can be passed around to different function calls for maximum
memory usage efficiency. This also makes it extremely easy to seamlessly
integrate \pkg{Moonshine} methods in a workflow that already takes advantage of
\pkg{Bumper.jl} facilities.

\subsection{Complete Algorithm}
The complete ARG inference procedure is given in \cref{alg:arg-construction}.

\begin{algorithm}
  \setstretch{1.05}
  \DontPrintSemicolon

  \SetKwData{Midx}{markeridx}
  \SetKwData{Mpos}{markerpos}
  \SetKwData{Ledges}{live\_edges}
  \SetKwData{Sc}{s\textsubscript{c}}
  \SetKwData{Dc}{d\textsubscript{c}}
  \SetKwData{Eone}{e\textsubscript{1}}
  \SetKwData{Etwo}{e\textsubscript{2}}
  \SetKwFunction{ItP}{idxtopos}
  \SetKwFunction{Parent}{parent}
  \SetKwFunction{Parents}{parents}
  \SetKwFunction{Src}{src}
  \SetKwFunction{Dst}{dst}
  \SetKwFunction{Sibling}{sibling}
  \SetKwData{Redge}{redge}
  \SetKwData{Rlat}{rlat}
  \SetKwData{Vupdate}{vupdate}
  \SetKwData{Cedge}{cedge}
  \SetKwData{Clat}{clat}
  \SetKwData{Pedges}{possible\_cedges}
  \SetKwData{K}{k}
  \SetKwData{Lbound}{lbound}
  \SetKwData{Ubound}{ubound}
  \SetKwFunction{Lat}{latitude}
  \SetKwData{E}{e}
  \SetKwData{R}{r}
  \SetKwArray{Mlats}{minlatitudes}
  \SetKwData{BP}{breakpoint}

  Set $\Midx \leftarrow 1$, $\Ledges \leftarrow \emptyset$ \;
  Update \Midx and \Ledges via \cref{alg:mmn} \;

  \BlankLine
  \While{$markeridx > 0$} {
    \While{$|\Ledges| > 1$} {
      Select \Eone \& \Etwo uniformly from \Ledges and remove them \;

      \BlankLine
      \tcp{Consider the possibility of a wild recombination}
      Set $\Mpos \leftarrow \ItP{\Midx}$ \;
      \uIf{$\Parent{\Src{\Eone}, \Mpos} = \Src{\Etwo}$} {
        Set $\Redge \leftarrow \Src{\Eone} - \Sibling{\Dst{\Eone}, \Mpos}$ \;
        Set $\Vupdate \leftarrow \Src{\Etwo}$ \;
      }
      \ElseIf{$\Parent{\Src{\Etwo}, \Mpos} = \Src{\Eone}$} {
        Set $\Redge \leftarrow \Src{\Etwo} - \Sibling{\Dst{\Etwo}, \Mpos}$ \;
        Set $\Vupdate \leftarrow \Src{\Eone}$ \;
      }
    }

    \BlankLine
    Select a recombination edge \Redge with probability proportional to its
      length among all admissible edges (see
      \cref{recombination-recoalescence} for admissibility criteria) \;
    Draw a recombination latitude \Rlat betwen \Dst{\Redge} and \Src{\Redge}
      from a position-scale family associated with the Beta(2, 2) distribution \;

    \BlankLine
    \uIf{Recombination is on a wild edge} {
      Draw a recoalescence latitude \Clat distributed as a non-homogeneous
        Poisson process on $[\Rlat, \infty)$ via empirical supremum rejection
        sampling (see \cref{recombination-recoalescence} for complete
        distribution) \;
      Select a recoalescence edge \Cedge uniformly among those intersecting
        latitude \Clat \;
    }
    \Else {
      Fill \Pedges with derived edges located above \Redge \;
      \For{$\K \in 1, \ldots, |\Pedges|$} {
        Set \R as to the latitude of the lowest valid recoalescence edge
          associated with \E according to \cref{recombination-recoalescence} \;
        Set $\Mlats \leftarrow \max \{ \Rlat, \R \}$ \;
      }

      Set $\Ubound \leftarrow \max \{ (\Lat \circ \Src)(\E) : \E \in \Pedges \}$ \;
      Set $\Lbound \leftarrow \min \Mlats $ \;
      Generate recoalescence latitude \Clat distributed as a non-homogeneous
        Poisson process on $[ \Lbound, \Ubound ]$ via rejection sampling \;
      Pick associated recoalescence edge \Cedge (see
        \cref{recombination-recoalescence} for complete procedure) \;
      Set \Vupdate to the number of vertices plus 2 \;
    }

    \BlankLine
    Choose \BP conditional on \Redge, \Cedge and \Midx (see \cref{breakpoint}) \;

    \BlankLine
    \uIf {$\Src{\Cedge} \in \Parents{\Dst{\Redge}}$ and \Dst{\Redge} is a
      recombination vertex} {
        Perform an MCO event on vertex \Dst{\Redge} \;
      }
    \Else {
      Generate a recombination event on \Redge at \Rlat followed by a
        recoalescence event on \Cedge at \Clat at position \BP \;
    }

    \BlankLine
    Update ARG via \cref{alg:arg-update} starting at \Vupdate \;
    Update \Ledges \;
    Update \Midx and \Ledges via \cref{alg:mmn} \;
  }

  \caption{ARG Inference}\label{alg:arg-construction}
\end{algorithm}

\section{Comparison With Other Methods}
\subsection{Number of Recombination Events}
\begin{figure}
  \centering
  \includegraphics[width = \columnwidth]{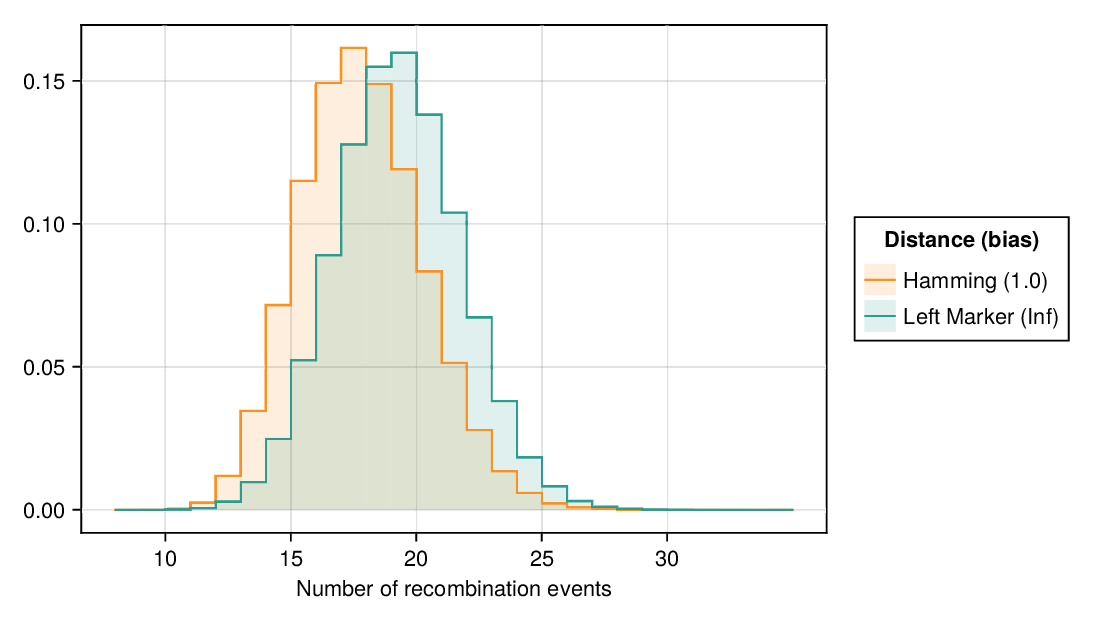}
  \caption{Distribution of the number of recombination events on ARGs inferred for the Kreitman dataset. The data are split into histograms based on the distance used to construct the initial coalescent tree.}\label{fig:recombination-kreitman}
\end{figure}

To study the distribution of the number of recombination events generated by \pkg{Moonshine}, we inferred two million ancestral recombination graphs for the well-known Kreitman dataset \cite{Kreitman1983}. As in \cite{Wong2024}, we followed the \pkg{stdpopsim} catalog \cite{Adrion2020,Lauterbur2023} and assumed a per-marker mutation rate and effective population size of $5.49 \time 10^{-9}$ and $1720600$, respectively. Half of these ARGs were based on trees constructed with the Hamming distance while the other half used the left marker distance. In both cases, the bias term $c_0 = 1$. Distributions are illustrated in \cref{fig:recombination-kreitman}.

The Hamming distance led to slightly more parsimonious ancestries. On average, they contained 17.43 (SD 2.52) recombination events versus 18.86 (SD 2.5) for the left marker-based counterpart. Additionally, the most parsimonious ancestry inferred with the Hamming distance contained 8 recombination events, just one above the theoretical minimum (assuming no recurrent mutation). This number was 9 for the left-marker-based ancestry. In both cases, the maximum number of recombination events inferred was 34.

We can compare these results with those obtained by \cite{Wong2024}. Most of the ARGs inferred by \pkg{Moonshine} were less parsimonious than the ancestry produced by \pkg{tsinfer}, which contained 11 breakpoints. \pkg{Relate} \cite{Speidel2019}, which allows for some recurrent mutations attributable to data error, is consistently more parsimonious with only 2 breakpoints. Note that neither of these two methods explicitly infers recombination events. \pkg{ARGWeaver} was able to construct an ARG with 37 recombination events using the SMC' model, more than any ARG produced by our package. Finally, \pkg{Moonshine} is not as parsimonious as \pkg{KwARG}, which is able to reach the theoretical minimum. Note that \pkg{KwARG} uses a heuristic specifically designed for parsimony.

\cite{Thao2019} provides additional comparisons. It introduces \pkg{GAMARG} which, similar to \pkg{KwARG}, infers ARGs via a parsimony heuristic and is able to reach the theoretical minimum. In addition, the paper reports a numerical experiment in which 1000 ancestries were inferred by \pkg{ARG4WG}, \pkg{REARG} \cite{PhuongThao2017} and \pkg{Margarita}. The first two attained a minimum of 10 events while the last one was on par with \pkg{Moonshine} with 8.

\subsection{Coalescence Times and Topology Accuracy}
To assess the quality of the ancestries reconstructed by \pkg{Moonshine}, we simulated a topology for a sample of 100 genetic sequences of length $10^5$, a recombination rate of $\rho = 10^{-8}$ and an effective population size of $10^4$. We then generated mutations on top of this topology according to the ISM using rates $\mu \in \{ \rho / 2, \rho, 2\rho \}$ yielding three ARGs with 119, 203 and 408 polymorphic sites, respectively. For each of these, we inferred 200 ancestries from the resulting haplotypes using the Hamming distance for the initial tree and an infinite window width. For comparison, we reconstructed ARGs from the same sample using \pkg{SINGER} and \pkg{Relate}.

\pkg{Relate} outputs a single ancestry and is straightforward to use. Conversion to \pkg{tskit}'s \code{TreeSequence} was performed via the included utility. ARG reconstruction was about 20 times slower than for \pkg{Moonshine} across mutation rates. Additionally, \pkg{Relate} failed to map 2 and 1 mutations for $\mu = 5 \times 10^{-9}$ and $\mu = 1 \times 10^{-8}$, respectively.

\pkg{SINGER} outputs multiple ARGs using a MCMC scheme and requires choosing a thinning interval as well as a burn-in period. For the former, we used the default of discarding 20 ancestries between samples. Under this regime, we sampled 2000 ARGs and assessed convergence, which was attained for every parameter of interest after about 1000 iterations. The last 200 ARGs were stored for analysis. We used the included \proglang{Python} script for conversion to \code{TreeSequence}. We had to manually reorder mutations due to a known bug of the script when dealing with flipped SNPS. Although we used the recommended prior probability of correct polarization of 0.99, some flips did occur, about 0.8 by ARG on average for $\mu = 2 \times 10^{-8}$. Reconstruction of a single ARG was between 58 and 21 times slower than for \pkg{Moonshine} with the difference decreasing as the mutation rate (and hence the number of markers) increases. Note that this does not account for ancestries discarded during the burn-in period.

\begin{figure}
  \centering
  \includegraphics[width = \columnwidth]{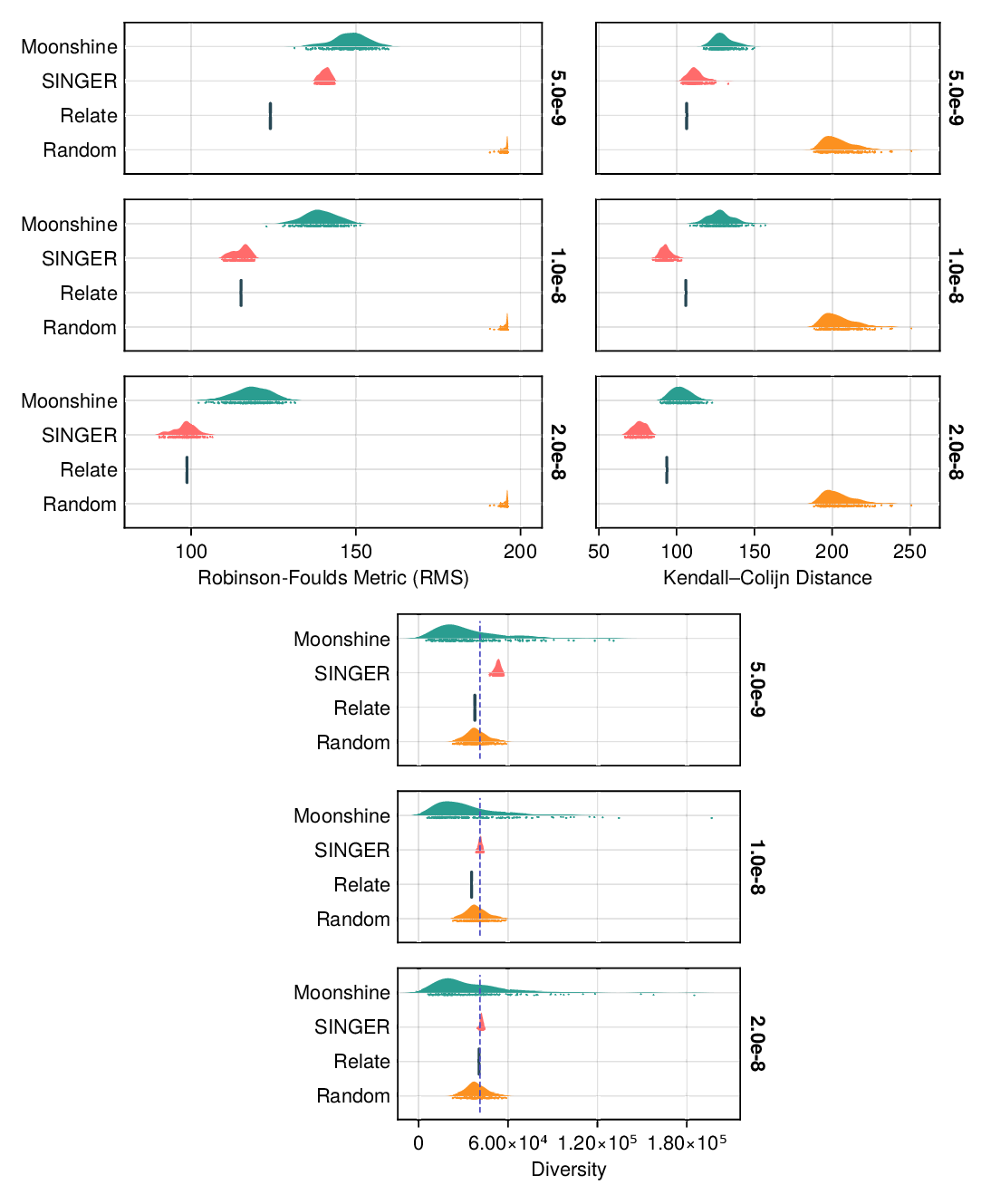}
  \caption{Accuracy of ARG reconstruction. Mutation rate is indicated on the right of the figures. An estimate of the density is provided for each metric for both \pkg{Moonshine} and \pkg{SINGER}. Since \pkg{Relate} is deterministic, the unique computed value is indicated by a vertical line. The dashed line in the bottom plot marks the true diversity of the simulated ARG.}\label{fig:recombination-time-topo}
\end{figure}

We compared ARGs using three metrics: the Robinson-Foulds (RF) distance \cite{Robinson1981}, the Kendall–Colijn (KC) distance \cite{Kendall2016} and the branch-length-based diversity (i.e. average pairwise coalescence times). The first two metrics measure accuracy in topology reconstruction while the last one evaluates inference of pairwise coalescence times. Since the RF metric is only defined for trees, we evaluated it on a regular grid of 25 positions and reported the root mean squared error (RMSE) with respect to the original ARG. Each of these quantities was measured using the corresponding methods in \pkg{tskit}. To establish a baseline, we compared 200 simulated ARGs with the original for each metric. We employed the same parameters as we did for the original ARG. Results are presented in \cref{fig:recombination-time-topo}. The baseline is the ``Random'' category.

\pkg{Moonshine} behaves as expected with respect to topological metrics. In both cases, reconstruction accuracy increases with the number of markers. The same is true for \pkg{SINGER} and \pkg{Relate}, which are slightly more accurate than \pkg{Moonshine}. The diversity of ARGs reconstructed by \pkg{Moonshine} and \pkg{Relate} is largely unaffected by the number of markers. In the same regard, performance of \pkg{SINGER} is generally very good and improves as the number of markers increases.

%% file: sections/conclusion.tex
We have described a fast and efficient algorithm for ancestral recombination graph inference and its Julia implementation via the \pkg{Moonshine} package. In addition to the exact procedure, an approximate scheme based on user-defined genomic windows is available, which reduces inference time. Inference is grounded in a restricted version of the coalescent-with-recombination distribution rather than heuristics. We have also presented a flexible algorithm for coalescent tree construction that can accommodate a wide range of metrics on haplotypes, exactly or approximately, while avoiding numerical instabilities.

One of the main drawbacks of \pkg{Moonshine} is the need for high-quality phased and polarized data. This limits the applicability of our method, as such data is relatively scarce in practice. Moreover, our approach is limited to binary markers and assumes non-recurrent mutations. Some of these limitations are more significant than others. For instance, the data structure used for storing haplotypes is fairly straightforward to adapt to multiallelic markers, and minor modifications to our algorithms would be sufficient to account for simple, non-recurrent mutation models, which mainly involve the existence of an absorbing state.

As of version 0.3.3, these algorithms are for the most part mature. No significant leap in inference time or memory usage is to be expected in the near future. Our goal for version 1.0.0 is to improve real-world usability. Specifically, the API reference needs to be completed to make implementing alternative models a more pleasant experience. General documentation and guides also need to be improved for the general user. Both are publicly available at \url{https://moonshine.patrickfournier.ca}.

As of today, version 0.3.4 and later are compatible with files encoded in the variant call format (VCF) through the \pkg{VCFTools.jl} package \citep{Chu2023}. We plan to support FASTA/FASTQ and sequence/binary alignment map format (SAM/BAM) via the \pkg{FASTX.jl} and \pkg{XAM.jl} packages, respectively. Along with compatibility with these input formats, we plan to support missing markers by treating them as non-ancestral material, which is the approach taken by \cite{Lyngsoe2005, Minichiello2006, Ignatieva2021}.

\pkg{Moonshine} is available on \proglang{Julia}'s general registry and can be easily and quickly obtained using standard facilities. In addition, we plan to package it in two application containers. One will include the Pluto notebook~\citep{FonsPlas2025} package and graphical utilities such as \pkg{Makie.jl} \citep{Danisch2021}. Its goal is to reduce the burden associated with ARG inference for practitioners. The other container will be more minimal, including only the minimum required to deploy \pkg{Moonshine} via an orchestration system such as \pkg{kubernetes}. Our hope is to facilitate the creation of high-performance computation clusters in small-to-medium environments.

%% file: sections/simulations.tex
\subsubsection{Results}
All simulations were performed with \pkg{Moonshine} version 0.3.9 and Julia 1.11.3 on AMD EPYC 9654 processors. Execution time was measured using the \code{@timed} macro. To reduce variability due to the execution environment and differences between ARGs, each reported measurement is the minimum of three runs, averaged over five graphs. Haplotypes are generated by \pkg{msprime} using \code{StandardCoalescent} and \code{BinaryMutationModel} as the ancestry and mutation models, respectively. Per-locus mutation and recombination rates were both set to $10^{-8}$. An effective population size of $10^4$ was specified. The calls to \code{msprime.sim\_ancestry} and \code{msprime.sim\_mutations} were made through the \pkg{Moonshine} interface to \pkg{msprime}.

All the code used for the simulation studies, as well as the raw data, is publicly available on Codeberg (\url{https://codeberg.org/ptrk/moonshine.jl-papers}). For convenient reproducibility, code to execute each simulation presented and produce related figures is grouped in a single Pluto notebook. As the simulations are computationally intensive and require considerable time to complete, it may be more convenient to run them on more powerful machines such as those found in a computer cluster. For that purpose, the notebook can be executed as a standalone Julia script. Figure generation steps will be skipped when doing so; they can be run locally later on.

Raw simulation results are presented below. Numbers of markers and recombination events are averaged over constructed ARGs.

\subimport{../assets}{tree-build-time-benchmark-table.tex}

\subimport{../assets}{tree-sampling-threshold-benchmark-table.tex}

\subimport{../assets}{arg-winwidth-benchmark-table.tex}

%% file: assets/tree-build-time-benchmark-table.tex
\begin{longtable}{cccccc}
\caption{Tree Building Time}\label{table:tree-build-time-benchmark}\\
\toprule
n & L [Mbp] & Distance & \# Markers & Time & Size [MByte]\\
\midrule
10 & 1 & Hamming & 588.4 & $< 1$s & $< 1$\\
10 & 1 & LeftM & 588.4 & $< 1$s & $< 1$\\
10 & 10 & Hamming & 5725.4 & $< 1$s & $< 1$\\
10 & 10 & LeftM & 5725.4 & $< 1$s & $< 1$\\
10 & 100 & Hamming & 56517.4 & $< 1$s & $< 1$\\
10 & 100 & LeftM & 56517.4 & $< 1$s & $< 1$\\
10 & 250 & Hamming & 141851 & $< 1$s & 1\\
10 & 250 & LeftM & 141851 & $< 1$s & 1\\
100 & 1 & Hamming & 1035.2 & $< 1$s & $< 1$\\
100 & 1 & LeftM & 1035.2 & $< 1$s & $< 1$\\
100 & 10 & Hamming & 10435.4 & $< 1$s & $< 1$\\
100 & 10 & LeftM & 10435.4 & $< 1$s & $< 1$\\
100 & 100 & Hamming & 103598 & $< 1$s & 3\\
100 & 100 & LeftM & 103598 & $< 1$s & 3\\
100 & 250 & Hamming & 259113 & $< 1$s & 8\\
100 & 250 & LeftM & 259113 & $< 1$s & 8\\
1000 & 1 & Hamming & 1539.8 & $< 1$s & $< 1$\\
1000 & 1 & LeftM & 1539.8 & $< 1$s & $< 1$\\
1000 & 10 & Hamming & 14950.2 & $< 1$s & 4\\
1000 & 10 & LeftM & 14950.2 & $< 1$s & 4\\
1000 & 100 & Hamming & 149768 & $< 1$s & 38\\
1000 & 100 & LeftM & 149768 & $< 1$s & 38\\
1000 & 250 & Hamming & 375008 & $< 1$s & 96\\
1000 & 250 & LeftM & 375008 & $< 1$s & 96\\
10000 & 1 & Hamming & 1928.2 & 00:00:04 & 6\\
10000 & 1 & LeftM & 1928.2 & 00:00:02 & 6\\
10000 & 10 & Hamming & 19569.2 & 00:00:15 & 50\\
10000 & 10 & LeftM & 19569.2 & 00:00:02 & 50\\
10000 & 100 & Hamming & 195747 & 00:01:11 & 492\\
10000 & 100 & LeftM & 195747 & 00:00:03 & 492\\
10000 & 250 & Hamming & 489204 & 00:02:12 & 1228\\
10000 & 250 & LeftM & 489204 & 00:00:03 & 1228\\
\bottomrule
\end{longtable}

Legend: \textbf{n}: number of haplotypes in the sample, \textbf{L}: length of haplotypes, \textbf{Distance}: haplotype distance used for construction, \textbf{\# Markers}: number of markers, \textbf{Time}: construction time, \textbf{Size}: memory usage of the tree.

%% file: assets/tree-sampling-threshold-benchmark-table.tex
\begin{longtable}{ccccc}
\caption{Tree Sampling Threshold}\label{table:tree-sampling-threshold-benchmark}\\
\toprule
Threshold & \# Markers & Time & Size [MByte] & SeU\\
\midrule
0.1 & 491025 & 00:00:09 & 1228 & 11.1\\
0.2 & 491025 & 00:00:21 & 1228 & 5\\
0.3 & 491025 & 00:00:32 & 1228 & 3.3\\
0.4 & 491025 & 00:00:43 & 1228 & 2.4\\
0.5 & 491025 & 00:00:55 & 1228 & 1.9\\
0.6 & 491025 & 00:01:01 & 1228 & 1.7\\
0.7 & 491025 & 00:01:13 & 1228 & 1.5\\
0.8 & 491025 & 00:01:30 & 1228 & 1.2\\
0.9 & 491025 & 00:01:38 & 1228 & 1.1\\
1 & 491025 & 00:01:46 & 1228 & ---\\
\bottomrule
\end{longtable}

Legend: \textbf{Threshold}: threshold for the secretary sampler, \textbf{\# Markers}: number of markers, \textbf{Time}: construction time, \textbf{Size}: memory usage of the tree, \textbf{SeU}: speedup. Speedups are computed with respect to a threshold of 1.

%% file: assets/arg-winwidth-benchmark-table.tex
\begin{longtable}{cccccccccc}
\caption{ARG Window Width}\label{table:arg-winwidth-benchmark}\\
\toprule
n & L [Mbp] & W [kbp] & \# Markers & \# Recs. & Time & Size [MByte] & SeU & SaU & RU\\
\midrule
10 & 1 & 0 & 549.2 & 175 & $< 1$s & $< 1$ & 1 & 1 & 1\\
10 & 1 & 100 & 549.2 & 173 & $< 1$s & $< 1$ & 0.8 & 1 & 1\\
10 & 1 & $\infty$ & 549.2 & 178 & $< 1$s & $< 1$ & --- & --- & ---\\
10 & 10 & 0 & 5694 & 1864 & $< 1$s & 1 & 1 & 1 & 1\\
10 & 10 & 100 & 5694 & 1843 & $< 1$s & 1 & 1.1 & 1 & 1\\
10 & 10 & $\infty$ & 5694 & 1784 & $< 1$s & 1 & --- & --- & ---\\
10 & 100 & 0 & 56502 & 19890 & 00:00:03 & 146 & 1.1 & 1 & 1\\
10 & 100 & 100 & 56502 & 19708 & 00:00:03 & 145 & 1 & 1 & 1\\
10 & 100 & $\infty$ & 56502 & 19661 & 00:00:04 & 146 & --- & --- & ---\\
10 & 250 & 0 & 141770 & 49793 & 00:00:22 & 903 & 1 & 1 & 1\\
10 & 250 & 100 & 141770 & 49432 & 00:00:21 & 901 & 1 & 1 & 1\\
10 & 250 & $\infty$ & 141770 & 49533 & 00:00:27 & 899 & --- & --- & ---\\
100 & 1 & 0 & 1064 & 1476 & $< 1$s & $< 1$ & 1.3 & 1 & 1\\
100 & 1 & 100 & 1064 & 1516 & $< 1$s & $< 1$ & 1.2 & 1 & 0.9\\
100 & 1 & $\infty$ & 1064 & 1417 & $< 1$s & $< 1$ & --- & --- & ---\\
100 & 10 & 0 & 10470 & 15452 & $< 1$s & 23 & 1.7 & 1 & 1\\
100 & 10 & 100 & 10470 & 15480 & 00:00:01 & 23 & 1 & 1 & 1\\
100 & 10 & $\infty$ & 10470 & 15413 & 00:00:01 & 23 & --- & --- & ---\\
100 & 100 & 0 & 103530 & 156060 & 00:00:35 & 2070 & 2 & 1 & 1\\
100 & 100 & 100 & 103530 & 155910 & 00:00:40 & 2068 & 1.8 & 1 & 1\\
100 & 100 & $\infty$ & 103530 & 154340 & 00:01:10 & 2046 & --- & --- & ---\\
100 & 250 & 0 & 259080 & 393250 & 00:06:17 & 12940 & 1.1 & 1 & 1\\
100 & 250 & 100 & 259080 & 390310 & 00:07:08 & 12883 & 1 & 1 & 1\\
100 & 250 & $\infty$ & 259080 & 388380 & 00:12:50 & 12780 & --- & --- & ---\\
1000 & 1 & 0 & 1473.8 & 5430 & $< 1$s & 2 & 1 & 1 & 1\\
1000 & 1 & 100 & 1473.8 & 5668 & $< 1$s & 2 & 1 & 1 & 1\\
1000 & 1 & $\infty$ & 1473.8 & 5423 & $< 1$s & 2 & --- & --- & ---\\
1000 & 10 & 0 & 15109 & 68810 & 00:00:08 & 151 & 1 & 1 & 1\\
1000 & 10 & 100 & 15109 & 66969 & 00:00:09 & 149 & 1 & 1 & 1\\
1000 & 10 & $\infty$ & 15109 & 64435 & 00:00:17 & 142 & --- & --- & ---\\
1000 & 100 & 0 & 149680 & 733560 & 00:04:50 & 13953 & 1.1 & 1 & 1\\
1000 & 100 & 100 & 149680 & 725930 & 00:05:14 & 13827 & 1 & 1 & 1\\
1000 & 100 & $\infty$ & 149680 & 695650 & 00:12:26 & 13254 & --- & --- & ---\\
1000 & 250 & 0 & 374300 & 1829300 & 01:09:57 & 86962 & 1 & 1 & 1\\
1000 & 250 & 100 & 374300 & 1822500 & 00:49:41 & 86482 & 1.4 & 1 & 1\\
1000 & 250 & $\infty$ & 374300 & 1747500 & 01:29:54 & 82919 & --- & --- & ---\\
10000 & 1 & 0 & 1941.4 & 14306 & 00:00:07 & 12 & 1 & 1 & 1\\
10000 & 1 & 100 & 1941.4 & 14605 & 00:00:06 & 12 & 1 & 1 & 1\\
10000 & 1 & $\infty$ & 1941.4 & 14505 & 00:00:07 & 12 & --- & --- & ---\\
10000 & 10 & 0 & 19805 & 247010 & 00:02:04 & 716 & 1.7 & 0.9 & 0.9\\
10000 & 10 & 100 & 19805 & 245310 & 00:02:06 & 707 & 1.6 & 0.9 & 0.9\\
10000 & 10 & $\infty$ & 19805 & 225920 & 00:03:27 & 659 & --- & --- & ---\\
10000 & 100 & 0 & 195590 & 3191500 & 01:26:41 & 79780 & 1 & 1 & 1\\
10000 & 100 & 100 & 195590 & 3154100 & 01:48:44 & 78914 & 0.8 & 1 & 1\\
10000 & 100 & $\infty$ & 195590 & 2728300 & 04:08:06 & 68569 & --- & --- & ---\\
10000 & 250 & 0 & 490510 & 8326400 & 05:58:34 & 518390 & 2.6 & 0.9 & 0.9\\
10000 & 250 & 100 & 490510 & 8220500 & 06:06:27 & 513540 & 2.5 & 0.9 & 0.9\\
10000 & 250 & $\infty$ & 490510 & 7149600 & 15:26:44 & 445150 & --- & --- & ---\\
\bottomrule
\end{longtable}

Legend: \textbf{n}: number of haplotypes in the sample, \textbf{L}: length of haplotypes, \textbf{W}: window width, \textbf{\# Markers}: number of markers, \textbf{\# Recs.}: number of recombination events, \textbf{Time}: construction time, \textbf{Size}: memory usage of the ARG, \textbf{SeU}: speedup, \textbf{SaU}: spaceup, \textbf{RU}: recup. Speedups, spaceups and recups are computed with respect to an infinite window.

%% file: sections/secretary-sampler.tex
\begingroup
\newcommand{\ks}{\widehat K}
\newcommand{\ktm}{K_{t_m}}
\newcommand{\spi}[2]{\sum_{i = #1}^{#2} p_i}

The following result gives the probability that a run of the secretary sampler
is exact.

\begin{lemma}
  \label{lemma:secretary}

  Let $K$ be the index of the sample returned by a run of the secretary sampler
  on a vector of size $m$ with threshold $t_0$, $t_m = \lfloor m t_0 \rfloor$,
  $p_i = p_{a b_i}$, and $\ktm = \argmax_{i = 1}^{t_m} \{ g_i + \log p_i \}$.
  Let $\ks$ be the index of the correct sample. Then,
  \begin{equation*}
    \Pr(K = \ks) = \frac{1}{\spi 1 m}
    \left( \spi{1}{t_m} +
    p_{\ktm}^{\ks - t_m - 1} \left( \spi{t_m + 1}{m} \right)
    \prod_{i = t_m + 1}^{\ks - 1} (p_{\ktm} + p_i)^{-1} \right)\,.
  \end{equation*}
\end{lemma}

\begin{proof}
  By the formula of total probability,
  \begin{equation*}
    \Pr(K = \ks) =
    \Pr(K = \ks \given \ks \leq t_m) \Pr(\ks \leq t_m) +
    \Pr(K = \ks \given \ks > t_m) \Pr(\ks > t_m)\,.
  \end{equation*}
  Since the algorithm visits every element with an index below $t_m$, $\Pr(K =
  \ks \given \ks \leq t_m) = 1$. Given that $\ks$ is distributed as a
  categorical random variable, we obtain
  \begin{align*}
    \Pr(\ks \leq t_m) = \frac{\spi{1}{t_m}}{\spi 1 m} &&
    \Pr(\ks > t_m) = \frac{\spi{t_m + 1}{m}}{\spi 1 m}\,.
  \end{align*}
  It only remains to show
  \begin{equation*}
    \Pr(K = \ks \given \ks > t_m) =
    p_{\ktm}^{\ks - t_m - 1} \prod_{i = t_m + 1}^{\ks - 1} (p_{\ktm} + p_i)^{-1}\,.
  \end{equation*}
  By pairwise independence of the sequence $g_1, \ldots, g_m$, we obtain,
  after grouping similar terms, the following expression:
  \begin{equation*}
    \Pr(K = \ks \given \ks > t_m) =
    \prod_{i = t_m + 1}^{\ks - 1} \Pr(g_i - g_{\ktm} \leq \log p_{\ktm} - \log p_i)
  \end{equation*}
  Since $g_i$ and $g_{\ktm}$ follow a standard Gumbel distribution, their
  difference is distributed as a standard Logistic distribution. In consequence,
  the probability of interest is nothing more than the CDF of this distribution.
  Routine simplifications complete the proof.
\end{proof}
\endgroup